\documentclass[reprint,superscriptaddress,showpacs,amsmath,amssymb,aps,pra]{revtex4-1}
\usepackage{amsthm}
\usepackage{graphicx}
\usepackage[ruled]{algorithm2e}
\usepackage{algorithmic}
\usepackage{hyperref}
\usepackage{appendix}
\usepackage{times}
\usepackage{color}

\hypersetup{colorlinks=true, citecolor=blue, urlcolor=blue, linkcolor=blue}
\begin{document}
\title{Improved iterative quantum algorithm for ground-state preparation}
\author{Jin-Min Liang}
\email{jmliang@cnu.edu.cn}
\affiliation{School of Mathematical Sciences, Capital Normal University, Beijing 100048, China}
\author{Qiao-Qiao Lv}
\affiliation{School of Mathematical Sciences, Capital Normal University, Beijing 100048, China}
\author{Shu-Qian Shen}
\affiliation{College of Science, China University of Petroleum, Qingdao 266580, China}
\author{Ming Li}
\affiliation{College of Science, China University of Petroleum, Qingdao 266580, China}
\author{Zhi-Xi Wang}
\email{wangzhx@cnu.edu.cn}
\affiliation{School of Mathematical Sciences, Capital Normal University, Beijing 100048, China}
\author{Shao-Ming Fei}
\email{feishm@cnu.edu.cn}
\affiliation{School of Mathematical Sciences, Capital Normal University, Beijing 100048, China}

\begin{abstract}
Finding the ground state of a Hamiltonian system is of great significance in many-body quantum physics and quantum chemistry. We propose an improved iterative quantum algorithm to prepare the ground state of a Hamiltonian. The crucial point is to optimize a cost function on the state space via the quantum gradient descent (QGD) implemented on quantum devices. We provide practical guideline on the selection of the learning rate in QGD by finding a fundamental upper bound and establishing a relationship between our algorithm and the first-order approximation of the imaginary time evolution. Furthermore, we adapt a variational quantum state preparation method as a subroutine to generate an ancillary state by utilizing only polylogarithmic quantum resources. The performance of our algorithm is demonstrated by numerical calculations of the deuteron molecule and Heisenberg model without and with noises. Compared with the existing algorithms, our approach has advantages including the higher success probability at each iteration, the measurement precision-independent sampling complexity, the lower gate complexity, and only quantum resources are required when the ancillary state is well prepared.

\end{abstract}

\maketitle

\section{Introduction}
One of the most promising applications of quantum computers is to simulate the dynamics of chemical and physical systems \cite{Feynman1982Simulating}. Quantum simulation in general requires to prepare an evolved state at any time and make measurement with respect to a physical observable \cite{Bolens2021Reinforcement,Lu2021Preparation,Lee2022Variational,LiangWeiFei2022Quantum,Xie2022Orthogonal}.
In particular, the ground state preparation for a given Hamiltonian system is of great significance. By using the Jordan-Wigner or Bravyi-Kitaev \cite{Bravyi2002} transformations, the molecule Hamiltonian can be transformed into qubit Hamiltonian in many quantum chemistry problems. The property of the ground states allows one to understand the dynamics of molecular and the design of drug \cite{Vivo2016Role}.

Many works have focused on the ground state preparation of a Hamiltonian \cite{Innocenti2020Ultrafast,Huggins2020A,Seki2021Quantum,Yao2021Reinforcement,Bierman2022Quantum}. For instance, the earlier work using quantum phase estimation prepares the ground state by projecting a guess state onto the ground state \cite{Abrams1999Quantum}. However, the long coherence time and high-fidelity gates render it impractical for noisy intermediate-scale quantum (NISQ) devices. Later, the variational quantum eigensolver (VQE) \cite{Peruzzo2014Variational} has attracted much attention due to its potential to be performed on NISQ devices. VQE obtains the ground state with a proper choice of ansatz, a suitable cost function and a controllable classical optimization \cite{Liang2020Variational,Cerezo2021Variational,Bharti2022Noisy,Callison2022Hybrid}. Nevertheless, VQE may face obstacles such as highly depending on the expressibility of the chosen ansatz and the so-called barren plateau which makes the optimization track detract from the global minimum \cite{McClean2018Barren}. In order to circumvent the barren plateau, imaginary time evolution (ITE) is an alternative which drives an initial state to the ground state after long time evolution \cite{Motta2020Determining,McArdle2019Variational,Gomes2021Adaptive,Lin2021Real}.

Apart from the above approaches, an iterative quantum technique utilizes the power iteration \cite{Kyriienko2020Quantum} and the linear combinational of unitary operators (LCU) \cite{Long2006General,Long2008Duality,Berry2015Simulating}. A representative work is the full quantum eigensolver (FQE) combining LCU and quantum gradient descent (QGD), which optimizes a cost function on original state space via iteratively performing gradient operator on an initial state \cite{Wei2020A}. This valuable technique has been generalized to approximate Hamiltonian operator \cite{Bespalova2021Hamiltonian}, optimize a polynomial function \cite{Li2021Optimizing}, determine the excited states of a Hamiltonian \cite{Wen2021A} and obtaine the generalized eigenstates of a matrix pencil \cite{Liang2022Quantum}.

Consider the time-independent Schr\"{o}dinger equation,
\begin{align}\label{EigenDecomp}
\hat{\mathcal{H}}|u_{i}\rangle=E_{i}|u_{i}\rangle,~~E_{0}<E_{1}\leq\cdots\leq E_{N-1},~~N=2^{n},
\end{align}
where $|u_{i}\rangle$ denotes the eigenstate of the Hamiltonian $\hat{\mathcal{H}}\in\mathbb{C}^{N\times N}$ with eigenvalue $E_{i}$. Assume that the $l$-local Hamiltonian encoded in an $n$-qubit system can be written as the linear combinations of unitary operators $\hat{\mathcal{H}}_{k}$,
\begin{align}\label{Hamiltonian}
\hat{\mathcal{H}}=\sum_{k=0}^{K-1}h_k\hat{\mathcal{H}}_k,
\end{align}
where $h_k$ are real coefficients and $\hat{\mathcal{H}}_k$ are tensor products of Pauli matrices $\{\sigma_{x},\sigma_{y},\sigma_{z}\}$ and the $2\times2$ identity matrix $I_{2}$. Assume also that the number of non-trivial nonzero terms being in $K\sim\mathcal{O}[\textrm{poly}(n)]$ which grows only polynomially with the number of qubits $n$.

The FQE prepares the ground state by optimizing a cost function $C(|\phi\rangle)=\langle\phi|\hat{\mathcal{H}}|\phi\rangle$ with the help of QGD and LCU \cite{Wei2020A}. Basically, the minimal value of $C(|\phi\rangle)$ implies the ground energy $E_{0}$ and ground state $|u_{0}\rangle$. The gradient of $C(|\phi\rangle)$ with respect to state $|\phi\rangle$ is $\nabla C(|\phi\rangle)=2\hat{\mathcal{H}}|\phi\rangle$. Applying the gradient descent formula, the update state after one iteration is given by
\begin{align}
|\phi(1)\rangle&=\frac{|\phi(0)\rangle-\mu\nabla C(|\phi(0)\rangle)}{\||\phi(0)\rangle-\mu\nabla C(|\phi(0)\rangle)\|_{2}}
=\frac{G|\phi(0)\rangle}{\|G|\phi(0)\rangle\|_{2}},
\end{align}
where $\mu>0$ denotes the learning rate and the non-unitary operator $G=I_{N}-2\mu\hat{\mathcal{H}}$. The iteration process can be seen as a pure state evolution under the non-unitary operator $G$. Although the implementation of operator $G$ gives rise to a challenge for quantum devices designed only from unitary gates, the dilation method \cite{Sweke2015Universal,Marsden2021Capturing} makes it possible by generating a superposition state whose elements are associated with the coefficients $h_k$. After carefully selecting the learning rate $\mu$ in practical situations, FQE converges to the ground state with polylogarithmic depth circuit. Moreover, it has been demonstrated that FQE can prepare the ground state even at the presence of Gaussian and random noise \cite{Wei2020A}.

However, there are three caveats which may hinder the efficient implementations of FQE on quantum computers. (i) The selection of learning rate $\mu$. Different learning rate may have different performance on the convergence. Thus, in practice the selection of the learning rate demands a theoretical constraint. (ii) The preparation of the ancillary superposition state. Before applying the gradient operator $G$, one needs to prepare an ancillary state whose elements are the coefficients of the Pauli decomposition of the gradient operator. For a general $2^{m}$-dimensional classical vector, it has been demonstrated that an exact universally algorithm would need at least $\mathcal{O}(m)$ qubits and $\mathcal{O}(2^{m})$ operators to prepare the corresponding amplitude encoding state \cite{Plesch2011Quantum}. Therefore, new technology is needed to reduce the gate complexity. (iii) The lack of analysis on the sampling complexity. Although quantum amplitude amplification (QAA) induces a quadratic speedup \cite{Brassard2002Quantum}, performing QAA needs additional computational resources.

In this work, we treat the above problems with several techniques and present an improved iterative quantum algorithm for ground state preparation. Our algorithm has a lower gate complexity attributed to the preparation of the ancillary state. For the problem (i), we theoretically demonstrate that the learning rate has an upper bound determined by the ground state energy and the largest eigenvalue of the Hamiltonian. In particular, the learning rate may be arbitrary positive number for special Hamiltonian. Moreover, we find that the first-order approximation of imaginary time evolution and our algorithm are equivalent. This claim provides a practical guideline about the selection of the learning rate. Specifically, $\mu=\Delta t/2$ for small time step $\Delta t$. For the problem (ii), we adapt a variational quantum state preparation (VQSP) to efficiently access an ancillary state with only polynomial overhead in terms of the size of the state. For the problem (iii), we demonstrate that the sampling complexity of our algorithm does not depend on the precision of measurement as shown in VQE. Actually, it is a finite value depending on the summation of the coefficients in the decomposition of gradient operator, the overlap between the initial state and the ground state, and the largest (in absolute value) eigenvalue of gradient operator. Meanwhile, the sampling complexity decreases with the increase of the iterations. Finally, we numerically prepare ground states of the deuteron molecule and Heisenberg model on noiseless and noisy situations. Since the behavior of Gaussian and random noises has been investigated in FQE \cite{Wei2020A}, here the noise behavior is modeled and simulated by accounting to the global depolarizing noise channel in each iteration. The fidelity and the evolved energy are chosen as the quantities to evaluate the performance of the obtained ground state. Comparing with VQE for the deuteron molecule, our algorithm has shorter iteration steps even the consumed resources are different. It does not need to perform classical-quantum optimization loops, thus avoiding the barren plateau as long as VQSP is perfectly achieved in advance. Finally, our algorithm has higher success probability at the end of each iteration compared with FQE scheme. Hence, the sampling complexity is reduced.

\section{Methods}\label{Sec:II}
\begin{figure}[ht]
\includegraphics[scale=0.5]{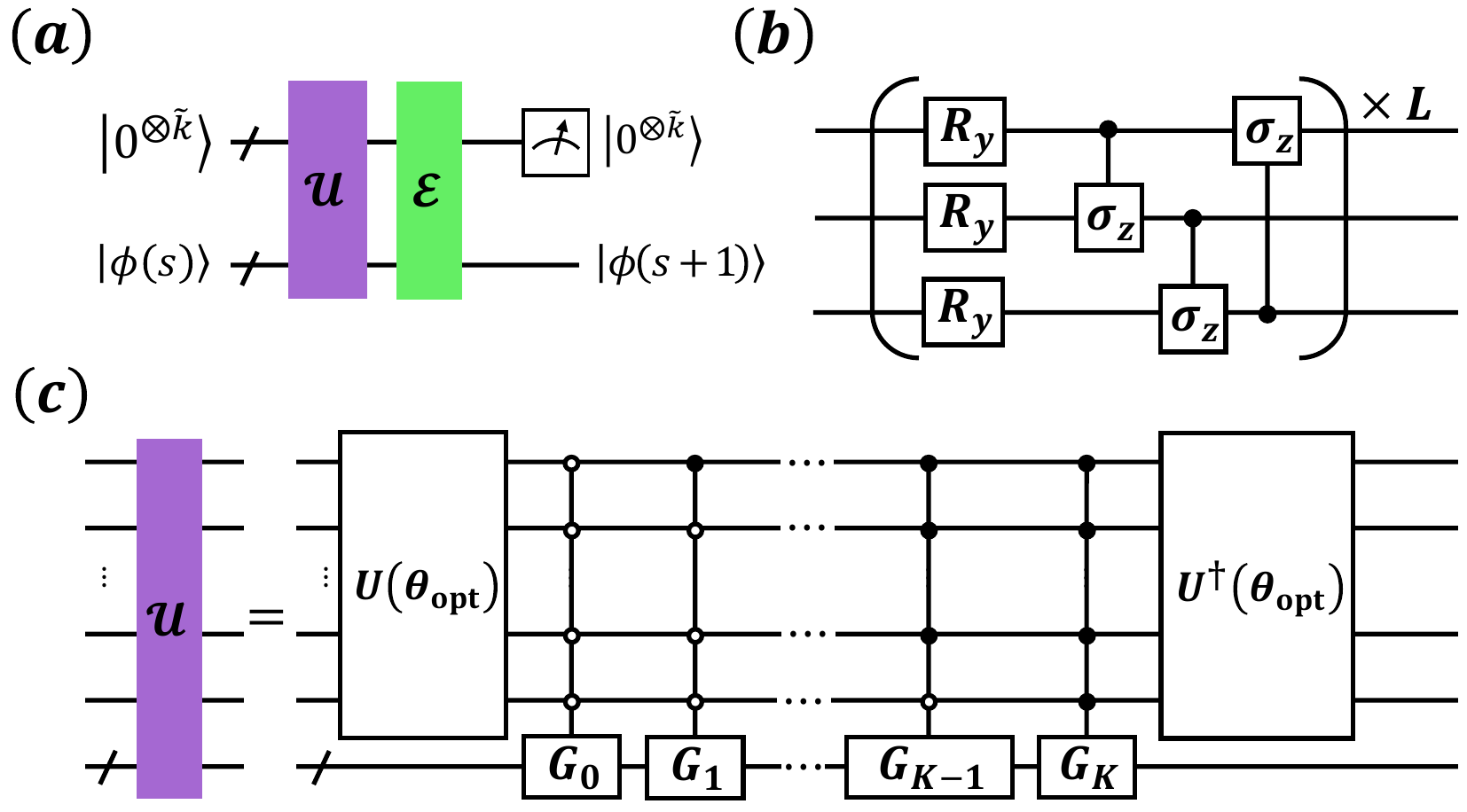}
\caption{(a) A schematic diagram of our iterative quantum algorithm for the ground state preparation. $\mathcal{E}$ denotes the noise channel. (b) A $3$-qubit parametrized quantum circuit $U(\boldsymbol\theta)$ in which the unitary $R_{y}$ is parameterized. (c) The decomposition of the unitary $\mathcal{U}=[U^{\dag}(\boldsymbol\theta_{\textrm{opt}})\otimes I_{N}]\Lambda_{\tilde{k}}G[U(\boldsymbol\theta_{\textrm{opt}})\otimes I_{N}]$.}
\label{Fig1schematic}
\end{figure}

Our iterative quantum algorithm for ground-state preparation contains two crucial subroutines: variational quantum state preparation (VQSP) presented in Sec. \ref{Sec:IIC:VQSP} and the implementation of the non-unitary gradient operator \cite{Berry2015Simulating}. A schematic diagram for the proposed algorithm is shown in Fig. (\ref{Fig1schematic}.a).

For a quantum system of Hamiltonian (\ref{Hamiltonian}), the gradient operator $G$ can also be expressed as an LCU,
\begin{align}\label{GradientOperator}
G&=I_{N}-2\mu\sum_{k=0}^{K-1}h_k\hat{\mathcal{H}}_k=\sum_{k=0}^{K}y_kG_k,~~y_k>0,
\end{align}
where the coefficients $y_k=|2\mu h_k|$ and the unitary $G_k=s_k\hat{\mathcal{H}}_k$ for $k=0,\cdots,K-1$, $y_K=1$ and $G_K=I_{N}$. Note that the sign $s_k=1$ $(-1)$ if $h_k$ is negative (positive). Here we assume that $K+1=2^{\tilde{k}}$ for an integer $\tilde{k}$. If $K+1$ is not a power of $2$, we can divide the identity operator $I_{N}$ into several sub-terms until $K+1$ can be denoted as a power of $2$.

Before executing our algorithm, we require to prepare a superposition state in a $\tilde{k}$-qubit system,
\begin{align}
|\boldsymbol{y}\rangle=\sum_{k=0}^{K}\boldsymbol{y}_k|k\rangle,
~~\boldsymbol{y}_k=\sqrt{\frac{y_k}{\mathcal{N}_{y}}},
\end{align}
where the constant $\mathcal{N}_{y}=\sum_{k=0}^{K}y_k$. We give a variational quantum algorithm to achieve the state preparation on an NISQ computer in Sec. \ref{Sec:IIC:VQSP}, where an optimal unitary $U(\boldsymbol{\theta}_{\textrm{opt}})$ determined by the parameter $\boldsymbol{\theta}_{\textrm{opt}}$ is utilized to produce the state $|\boldsymbol{y}\rangle=U(\boldsymbol{\theta}_{\textrm{opt}})|0^{\otimes\tilde{k}}\rangle$.

\subsection{The main procedure}\label{Sec:IIA:MainAlgorithm}
The inputs in our algorithm are the Hamiltonian $\hat{\mathcal{H}}=\sum_{k=0}^{K-1}h_k\hat{\mathcal{H}}_k$, a tensor product state $|0^{\otimes\tilde{k}}\rangle$ and the initial state $|\phi(0)\rangle$. A well-decided initial state $|\phi(0)\rangle$ should have a sufficient overlap with the ground state $|u_{0}\rangle$ such that $c_{0}^{(0)}=|\langle u_{0}|\phi(0)\rangle|=\mathcal{O}[\textrm{poly}(1/n)]$ \cite{He2021Quantum,Note1}. Denote $S$ the largest iterative step. Our algorithm iteratively run the following procedures for $s=0,1,\cdots,S-1$.

(a) Performing the unitary operator $U(\boldsymbol\theta_{\textrm{opt}})\otimes I_{N}$ on the state $|\Psi(s)\rangle=|0^{\otimes\tilde{k}}\rangle|\phi(s)\rangle$, we obtain the state $|\Psi_{a}(s)\rangle=|\boldsymbol{y}\rangle|\phi(s)\rangle$.

(b) Apply the controlled unitary $\Lambda_{\tilde{k}}G
=\sum_{k=0}^{K}|k\rangle\langle k|\otimes G_k$ on the whole system. The state $|\Psi_{a}(s)\rangle$ is transformed into the state
\begin{align}
|\Psi_{b}(s)\rangle=\sum_{k=0}^{K}\boldsymbol{y}_k|k\rangle\otimes G_k|\phi(s)\rangle.\nonumber
\end{align}

(c) Implementing the unitary $U^{\dag}(\boldsymbol\theta_{\textrm{opt}})\otimes I_{N}$ on the state $|\Psi_{b}(s)\rangle$, we obtain
\begin{align}
|\Psi_{c}(s)\rangle=|0^{\otimes\tilde{k}}\rangle
\sum_{k=0}^{K}\boldsymbol{y}^{2}_kG_k|\phi(s)\rangle
+|\Psi_{\perp}(s)\rangle,\nonumber
\end{align}
where the ancillary state of $|\Psi_{\perp}(s)\rangle$ is orthogonal to $|0\cdots0\rangle$.

(d) Measure the ancillary system with the projector $M=|0^{\otimes\tilde{k}}\rangle\langle0^{\otimes\tilde{k}}|\otimes I_{N}$. The state of the whole system after the measurement is
\begin{align}
\frac{M|\Psi_{c}(s)\rangle}{\sqrt{P(s+1)}}
&=\frac{|0^{\otimes\tilde{k}}\rangle\sum_{k=0}^{K}y_kG_k|\phi(s)\rangle}{\mathcal{N}_{y}\sqrt{P(s+1)}}\nonumber\\
&=|0^{\otimes\tilde{k}}\rangle|\phi(s+1)\rangle=|\Psi(s+1)\rangle\nonumber.
\end{align}
The success probability of preparing the state $|\phi(s+1)\rangle$ is
\begin{align}
P(s+1)&=\langle\Psi_{c}(s)|M^{\dag}M|\Psi_{c}(s)\rangle=\frac{\|G|\phi(s)\rangle\|_{2}^{2}}{\mathcal{N}_{y}^{2}}.\nonumber
\end{align}

Furthermore, the following Lemma gives a critical property of the success probability that $P(s+1)$ would not decay exponentially to zero with the iteration (see the proof in Appendix \ref{AppendixA}).
\newtheorem{lemma}{Lemma}
\begin{lemma}\label{Lemma1}
The success probability of each iteration is an increasing sequence such that
\begin{align}
P(1)\leq P(2)\leq\cdots\leq P(S).
\end{align}
\end{lemma}

Although QAA provides a quadratic speedup on the measurement sampling complexity, $\Theta[P^{-1/2}(s+1)]$, this step is expensive since the Hamiltonian simulation has a sophisticated circuit structure and requires many copies of pure states $|\Psi_{c}(s)\rangle\langle\Psi_{c}(s)|$ \cite{Llyod2014Quantum}. Instead, we perform the classical Monte Carlo sampling \cite{Montanaro2015Quantum} and as a result it needs roughly $\Theta[P^{-1}(s+1)]$ copies of $|\Psi_{c}(s)\rangle$ to generate the state $|\phi(s+1)\rangle$.

We consider two common metrics for the solution quality, the energy and fidelity. The energy is given as an expectation value of the Hamiltonian $\hat{\mathcal{H}}$ on the evolved state $|\phi(s)\rangle$ such that
\begin{align}
E(s)&=\langle\phi(s)|\hat{\mathcal{H}}|\phi(s)\rangle
=\sum_{k=0}^{K-1}h_k\langle\phi(s)|\hat{\mathcal{H}}_{k}|\phi(s)\rangle.
\end{align}
The largest step $S$ is determined by the convergence criterion such as the energy difference between two consecutive steps, $|E(s)-E(s-1)|\leq\varepsilon$, where $\varepsilon$ is an error threshold \cite{Gomes2021Adaptive}. The fidelity between two the pure states $|\phi(s)\rangle$ and $|u_{0}\rangle$ is defined as $F(|\phi(s)\rangle,|u_0\rangle)=\sqrt{|\langle\phi(s)|u_0\rangle|^2}$ \cite{Nielsen2000Quantum}.

\subsection{The implementation of the block diagonal unitary}\label{Sec:IIB:Unitary}
In order to enact the non-unitary operator $G$ (Eq. \ref{GradientOperator}) on an arbitrary state $|\phi(s)\rangle$, we consider the implementation of the unitary operation $\Lambda_{\tilde{k}}G=\sum_{k=0}^{K}|k\rangle\langle k|\otimes G_k$ in a similar way adopted in \cite{Barenco1995Elementary,Wei2020A,Liang2022Quantum}. Since $\Lambda_{\tilde{k}}G$ is a block diagonal unitary, it can be implemented by applying $K+1$ $\tilde{k}$-fold controlled $n$-qubit gates, $\Lambda_{\tilde{k}}G=\prod_{k=0}^{K}\mathcal{C}_{\tilde{k},|k\rangle}G_k$, where the unitary
\begin{align}
\mathcal{C}_{\tilde{k},|k\rangle}G_k=|k\rangle\langle k|\otimes G_k+\sum_{\substack{k'=0\\k'\neq k}}^{K}|k'\rangle\langle k'|\otimes I_{N}.
\end{align}
The index $\tilde{k}$ in $\mathcal{C}_{\tilde{k},|k\rangle}G_k$ denotes the number of control qubits, see Fig. (\ref{Fig1schematic}.c). The index $|k\rangle$ in $\mathcal{C}_{\tilde{k},|k\rangle}G_k$ denotes the controlled state. Since the Hamiltonian $\hat{\mathcal{H}}$ is $l$-local, each operator $\mathcal{C}_{\tilde{k},|k\rangle}G_k$ can then be decomposed into at most $l$ $\tilde{k}$-fold controlled single-qubit gates $\mathcal{C}_{\tilde{k},|k\rangle}\sigma_{a}^{(b)}$, where $a\in\{x,y,z\}$, $\sigma_{a}^{(b)}$ denotes the Pauli operator $\sigma_{a}$ acting on $b$th qubit of the working system, and the factor $s_k$ has been absorbed in one of the operator $\sigma_{a}^{(b)}$. Notice that if $s_k=-1$, the operator $s_k\sigma_{a}^{(b)}$ can be achieved by the unitary $R_{x}(2\pi)\sigma_{a}^{(b)}$ associated with a rotation operator with respect to the $x$ axe.

For an arbitrary unitary $\mathcal{P}\in \textrm{U}(2)$, the controlled operator
\begin{align}
\mathcal{C}_{\tilde{k},|K\rangle}\mathcal{P}=|K\rangle\langle K|\otimes\mathcal{P}+\sum_{k=0}^{K-1}|k\rangle\langle k|\otimes I_{2}
\end{align}
can be decomposed into three controlled unitary operators $\mathcal{C}_{\tilde{k}-1,|(K-1)/2\rangle}Q$, $\mathcal{C}_{1,|1\rangle}Q$ and $\mathcal{C}_{1,|1\rangle}Q^{\dag}$, as well as two Toffolli gates over $\tilde{k}-1$ qubits, where $Q^2=\mathcal{P}$ \cite{Bergholm2005Quantum}. The cost of simulating the two Toffolli gates is $\mathcal{O}(\tilde{k})$ which counts the number of single qubit and CNOT gates \cite{Xin2017Quantum}. For an arbitrary unitary $\mathcal{C}_{\tilde{k},|k\rangle}\mathcal{P}$, $k=0,\cdots,K$, we utilize the Pauli operator $\sigma_{x}$ to realize the transformation between the operators $\mathcal{C}_{\tilde{k},|k\rangle}\mathcal{P}$ and $\mathcal{C}_{\tilde{k},|K\rangle}\mathcal{P}$. Particularly, given the binary representation $k=k_{1}k_{2}\cdots k_{\tilde{k}}$, we have
\begin{align}
\mathcal{C}_{\tilde{k},|k\rangle}\mathcal{P}&=|k\rangle\langle k|\otimes\mathcal{P}
+\sum_{\substack{k'=0\\k'\neq k}}^{K}|k'\rangle\langle k'|\otimes I_{2}\nonumber\\
&=[\sigma_{x}]\mathcal{C}_{\tilde{k},|K\rangle}\mathcal{P}[\sigma_{x}]^{\dag},
\end{align}
where $[\sigma_{x}]=\sigma_{x}^{!k_1}\otimes\sigma_{x}^{!k_2}
\otimes\cdots\sigma_{x}^{!k_{\tilde{k}}}\otimes I_{2}$ with $!k_{i}$ representing the NOT operator that returns $1$ when $k_{i}=0$ and returns $0$ when $k_{i}=1$. Denoting $\mathcal{T}$ the cost of simulating the unitary $\mathcal{C}_{\tilde{k},|k\rangle}\mathcal{P}$, we have $\mathcal{T}=\mathcal{O}(\tilde{k}^{2})$. Thus, the total gate complexity of simulating $\Lambda_{\tilde{k}}G$ is roughly $\mathcal{T}_{\textrm{total}}=l(K+1)\mathcal{T}=\mathcal{O}[l(K+1)\widetilde{k}^{2}]$.

\subsection{Variational quantum state preparation}\label{Sec:IIC:VQSP}
Algorithm \ref{VQSP} presents the detailed process of variational quantum algorithm to prepare the state $|\boldsymbol{y}\rangle$.
\begin{algorithm}
\caption{Variational quantum state preparation}
\label{VQSP}
\textbf{Input:} a reference quantum state $|0^{\otimes\tilde{k}}\rangle$, a parameterized quantum circuit (PQC) $U(\boldsymbol\theta)$, and a desired precision $\varepsilon^{'}$.

(1) Randomly choose an initial parameters $\boldsymbol\theta$ and measure the parameterized quantum state $|\psi(\boldsymbol\theta)\rangle=U(\boldsymbol\theta)|0^{\otimes\tilde{k}}\rangle$ in the standard basis $\{|z\rangle\}$, where $z\in\{0,1,\cdots,K\}$. The probability of measurement outcome $z$ is denoted as $\boldsymbol{p}_{z}$.

(2) Estimate the cost function $F(\boldsymbol\theta)$ on an NISQ device.

(3) Train the PQC $U(\boldsymbol\theta)$ and find the optimal parameter $\boldsymbol\theta_{\textrm{opt}}$ under the termination condition $\emph{Cond}:F(\boldsymbol\theta)\leq\varepsilon^{'}$.

\textbf{Output:} the optimal parameter $\boldsymbol\theta_{\textrm{opt}}$ and the encoded state $|\boldsymbol{y}\rangle\approx|\psi(\boldsymbol\theta_{\textrm{opt}})\rangle
=U(\boldsymbol\theta_{\textrm{opt}})|0^{\otimes\tilde{k}}\rangle$.

\end{algorithm}

In step (1), the parameterized quantum circuit $U(\boldsymbol\theta)$ is a hardware-efficient ansatz \cite{Kandala2017Hardware,Havlivcek2019Supervised,Commeau2020Variational}. For instance, As shown in Fig. (\ref{Fig1schematic}b), the PQC consists of single qubit gates $R_{y}(\theta)=e^{-\iota\frac{\theta}{2}\sigma_{y}}$ depending on the tunable parameter $\theta$, the Pauli operator $\sigma_{y}$ and the controlled Pauli gates $\sigma_{z}$ applying to the neighbor qubits. The probability $\boldsymbol{p}_{z}$ is generated by measuring the given trial state multiple times. Let the total number of samples be $N_{\textrm{sample}}$ and the number of samples of state $|z\rangle$ be $N_{z}$. Thus the probability is $\boldsymbol{p}_{z}=N_{z}/N_{\textrm{sample}}$.

In step (2), we define a cost function
\begin{align}\label{CostFun}
F(\boldsymbol\theta)=F_{\textrm{KL}}(\boldsymbol\theta)
+\Bigg|\sum_{k=0}^{K}\boldsymbol{y}_{k}-\sqrt{K+1}
\langle+^{\otimes\tilde{k}}|\psi(\boldsymbol\theta)\rangle\Bigg|,
\end{align}
where $|+\rangle=(|0\rangle+|1\rangle)/\sqrt{2}$ and $|\cdot|$ denotes absolute value. The first term
\begin{align}
F_{\textrm{KL}}(\boldsymbol\theta)=-\sum_{k=0}^{K}
\boldsymbol{y}_{k}^{2}\log\frac{\boldsymbol{p}_{k}}{\boldsymbol{y}_{k}^{2}}
\end{align}
is dubbed as the Kullback-Leibler divergence (KL) which quantifies the amount of information lost when changing from the vector distribution $\boldsymbol{y}^{[2]}=(\boldsymbol{y}_{0}^{2},\cdots,\boldsymbol{y}_{K}^{2})$ to another distribution $\boldsymbol{p}=(\boldsymbol{p}_{0},\cdots,\boldsymbol{p}_{K})$. The second term ensures the obtained states learned by the quantity $F_{\textrm{KL}}(\boldsymbol\theta)$ have positive local phases. For example, we aim to prepare a single qubit state $|\boldsymbol{y}\rangle=\frac{1}{\sqrt{5}}|0\rangle+\frac{2}{\sqrt{5}}|1\rangle$ corresponding to a classical vector $\boldsymbol{y}=(\frac{1}{\sqrt{5}},\frac{2}{\sqrt{5}})$. Variational optimizing the cost function $F_{\textrm{KL}}$ generates four states,
\begin{align}
&|\boldsymbol{y}_{0}\rangle=\frac{1}{\sqrt{5}}|0\rangle+\frac{2}{\sqrt{5}}|1\rangle,~
|\boldsymbol{y}_{1}\rangle=-\frac{1}{\sqrt{5}}|0\rangle+\frac{2}{\sqrt{5}}|1\rangle,\nonumber\\
&|\boldsymbol{y}_{2}\rangle=\frac{1}{\sqrt{5}}|0\rangle-\frac{2}{\sqrt{5}}|1\rangle,~
|\boldsymbol{y}_{3}\rangle=-\frac{1}{\sqrt{5}}|0\rangle-\frac{2}{\sqrt{5}}|1\rangle.\nonumber
\end{align}
The desired state $|\boldsymbol{y}_{0}\rangle$ can be obtained by optimizing the second term. Note that the second term can also be
\begin{align}
\left(\sum_{k=0}^{K}\boldsymbol{y}_{k}-\sqrt{K+1}\langle+^{\otimes\tilde{k}}|\psi(\boldsymbol\theta)\rangle\right)^2.
\end{align}

It is clear that $F(\boldsymbol\theta)$ is positive and is zero if and only if $U(\boldsymbol\theta_{\textrm{opt}})|0^{\otimes\tilde{k}}\rangle=|\boldsymbol{y}\rangle$. Much small cost values indicate a high fidelity preparation of the state $|\boldsymbol{y}\rangle$. Thus, the cost function is faithful and operationally meaningful \cite{Cerezo2021Variational}. Here the inner product $\langle+^{\otimes\tilde{k}}|\psi(\boldsymbol\theta)\rangle$ is computed via the Hadamard test \cite{Aharonov2009A}. Updating the circuit parameter by performing a classical optimization procedure such as the Nelder-Mead algorithm \cite{Nelder1965A}, we produce the optimal parameter $\boldsymbol\theta_{\textrm{opt}}$ until the cost function converges. The learning scheme maps the classical vector into a set of parameters $\boldsymbol\theta_{\textrm{opt}}$ such that $\boldsymbol{y}\mapsto\{\boldsymbol\theta_{\textrm{opt}}\}$. Loading the parameter $\boldsymbol\theta_{\textrm{opt}}$ into NISQ devices equipped with a parameterized quantum circuit $U(\boldsymbol\theta)$, we then produce the state $|\boldsymbol{y}\rangle\approx|\tilde{\boldsymbol{y}}\rangle
=U(\boldsymbol\theta_{\textrm{opt}})|0^{\otimes\tilde{k}}\rangle$.

\subsection{The selection of the learning rate}\label{Sec:IID:Equivalence}
In this section, we report strategies to determine the learning rate $\mu$. Theorem \ref{Theorem1} provides a theoretical benchmark for analyzing the learning rate $\mu$ (the proof is provided in Appendix \ref{AppendixB}).

\newtheorem{theorem}{Theorem}[]
\begin{theorem}[]\label{Theorem1}
Consider a system Hamiltonian $\hat{\mathcal{H}}$ (\ref{EigenDecomp}) with a ground state $|u_{0}\rangle$ and the ground state energy $E_{0}$. The algorithm proposed in Sec. \ref{Sec:IIA:MainAlgorithm} converges to the ground state $|u_{0}\rangle$ if the learning rate $\mu$ in Eq. (\ref{GradientOperator}) satisfies
\begin{equation}\label{ConditionLearningRate}
\mu\in\left\{
\begin{aligned}
&\Big(0,\frac{1}{E_{N-1}+E_0}\Big),&E_{N-1}+E_0>0\\
&\mathbb{R}^{+},&\textrm{others}\\
\end{aligned},
\right.
\end{equation}
where $\mathbb{R}^{+}$ denotes the set of positive real value and $E_{N-1}$ is the largest eigenvalue of $\hat{\mathcal{H}}$.
\end{theorem}

We here remark that from Theorem 1 the learning rate $\mu$ can be an arbitrary positive number when the eigenvalues satisfy $E_{N-1}+E_0\leq0$. One may wonder whether one can add diagonal constant shift to the Hamiltonian $\hat{\mathcal{H}}$ to switch the condition $E_{N-1}+E_0\leq0$.
Let $\hat{\mathcal{H}}^{'}=\hat{\mathcal{H}}+\tau I_N$ with eigenvalues $E_i^{'}=E_i+\tau$ for some constant $\tau\in\mathbb{R}$, such that $E_{N-1}^{'}+E_0^{'}\leq0$ but $E_{N-1}+E_0>0$. Based on Theorem \ref{Theorem1}, the corresponding gradient operator is given by $G^{'}=I_N-2\mu^{'}\hat{\mathcal{H}}^{'}$ with an arbitrary positive valued learn rate $\mu^{'}$. However, $E_{N-1}^{'}+E_0^{'}\leq0$ means that the shift constant $\tau\leq-\frac{E_{N-1}+E_0}{2}$. Therefore, although appropriate shift constant can avoid the selection of the learning rate, the fundamental convergence behavior does not modify.

We also remark that the condition (\ref{ConditionLearningRate}) is impractical since the eigenvalues of $\hat{\mathcal{H}}$ are required to be estimated in advance. Thus, we next provide a practical strategy to choose $\mu$ by demonstrating the equivalence between the imaginary time evolution (ITE) and the algorithm proposed in Sec. \ref{Sec:IIA:MainAlgorithm}.

ITE is an iterative computational method to solve the ground-state of many-body quantum systems \cite{Aulicino2021State}. Consider the time-independent Schr\"{o}dinger equation in imaginary time ($t\rightarrow-\iota t$),
\begin{align}\label{ITE}
\frac{\partial|\phi(t)\rangle}{\partial t}=-\hat{\mathcal{H}}|\phi(t)\rangle,
\end{align}
where $|\phi(t)\rangle$ is a quantum state at time $t$ and $\hat{\mathcal{H}}$ is the Hamiltonian. The formal solution of Eq. (\ref{ITE}) can be expressed as $|\phi(t)\rangle=e^{-\hat{\mathcal{H}}t}|\phi(0)\rangle$, where $|\phi(0)\rangle$ is an initial state at time $t=0$. Suppose the quantum state $|\phi(0)\rangle$ is expanded in the eigenbasis of $\hat{\mathcal{H}}$ given in (\ref{EigenDecomp}),
\begin{align}
|\phi(0)\rangle=\sum_{i=0}^{N-1}c_{i}^{(0)}|u_{i}\rangle,~c_{i}^{(0)}=\langle u_{i}|\phi(0)\rangle,~\sum_{i=0}^{N-1}|c_{i}^{(0)}|^2=1.\nonumber
\end{align}
The (unnormalized) sequence states are given by
\begin{align}
|\phi(t)\rangle=\sum_{i=0}^{N-1}c_{i}^{(0)}e^{-E_{i}t}|u_{i}\rangle\approx c_{0}^{(0)}e^{-E_{0}t}|u_{0}\rangle\nonumber
\end{align}
in the limitation of large $t$ with non-zero overlap $(c_{0}^{(0)}\neq0)$ \cite{Lehtovaara2006Solution}. The trial energy
\begin{align}
\lim_{t\rightarrow\infty}E(t)=\lim_{t\rightarrow\infty}
\frac{\langle\phi(t)|\hat{\mathcal{H}}|\phi(t)\rangle}{\langle\phi(t)|\phi(t)\rangle}=E_{0}.
\end{align}
Namely, the ITE always converges to the ground state after long time iterations.

Consider a small time $\Delta t=t/N_{t}$. The exponential operator $e^{-\hat{\mathcal{H}}t}=e^{-\hat{\mathcal{H}}N_{t}\Delta t}$ and each non-unitary operator $e^{-\hat{\mathcal{H}}\Delta t}$ have good approximations in terms of the following $J$th-order product formula \cite{Childs2021Theory},
\begin{align}
e^{-\hat{\mathcal{H}}\Delta t}=\sum_{j=0}^{J}\frac{(-\Delta t)^{j}}{j!}\hat{\mathcal{H}}^{j}+\mathcal{O}(\Delta t^{J+1}),
\end{align}
where the Taylor series is truncated at order $J$ \cite{Berry2015Simulating}. In particular, with the first-order truncation of the Taylor series, $e^{-\hat{\mathcal{H}}\Delta t}$ is approximated as a non-unitary operator
\begin{align}\label{ITMGradient}
\tilde{G}=I_{N}-\hat{\mathcal{H}}\Delta t,
\end{align}
with error $\mathcal{O}(\Delta t^{2})$. By representing $\tilde{G}$ as an LCU, the non-unitary operation $\tilde{G}$ can be implemented on a quantum computer \cite{Berry2015Simulating,Long2006General}. We repeatedly apply the non-unitary operator $\tilde{G}$ on a random initial state $|\phi(0)\rangle$. When the time $t$ is large enough, the updated state can be seen as an approximate ground state.

We observe that the two operators
$\tilde{G}=I_{N}-\Delta t\hat{\mathcal{H}}$ and $G=I_{N}-2\mu\hat{\mathcal{H}}$ are equivalent
when the small time step $\Delta t=2\mu$. Thus, in practice we select the learning rate $\mu$ in terms of $\Delta t$ such that $\mu=\Delta t/2$. $\Delta t$ should be sufficient small. Fox example, let the error of the first-order approximation of operator $e^{-\hat{\mathcal{H}}\Delta t}$ be $\epsilon=\Delta t^2$. This means that we should select $\mu=\Delta t/2=\sqrt{\epsilon}/2$. In the next section, we find that $\epsilon\leq10^{-2}$ is an amenable error.

\section{Numerical results and discussion}
\subsection{Numerical Simulation Results}
In this subsection, we apply the proposed algorithm to simulate the ground state of two models including the deuteron molecule \cite{Dumitrescu2018Cloud,Shehab2019Toward,Aydeniz2020Practical} and the Heisenberg spin$-1/2$ model with external magnetic field \cite{Bespalova2021Hamiltonian}.

\subsubsection{The deuteron molecule}
The $n=2$ oscillator-basis deuteron Hamiltonian in the discrete variable representation using the harmonic oscillator basis can be written as a Pauli string form,
\begin{align}\label{DMhamiltonian}
\hat{\mathcal{H}}&=5.907I_{4}+0.2183\sigma_{z}^{(1)}-6.125\sigma_{z}^{(2)}\nonumber\\
&-2.143(\sigma_{x}^{(1)}\sigma_{x}^{(2)}+\sigma_{y}^{(1)}\sigma_{y}^{(2)}),
\end{align}
where the Pauli operators $\sigma_{x}^{(k)}$, $\sigma_{y}^{(k)}$ and $\sigma_{z}^{(k)}$ act on the $k$th site. Based on the decomposition of $\hat{\mathcal{H}}$ in Eq. (\ref{DMhamiltonian}), the Pauli decomposition of the gradient operator $G$ is given by
\begin{align}
G&=I_{4}-2\mu\hat{\mathcal{H}}\nonumber\\
&=0.2I_{4}+0.3I_{4}+0.5I_{4}+11.814\mu R_{x}^{(1)}(2\pi)I_{4}\nonumber\\
&+0.4366\mu R_{x}^{(1)}(2\pi)\sigma_{z}^{(1)}+12.25\mu\sigma_{z}^{(2)}+4.286\mu\sigma_{x}^{(1)}\sigma_{x}^{(2)}\nonumber\\
&+4.286\mu\sigma_{y}^{(1)}\sigma_{y}^{(2)}.\nonumber
\end{align}
We here divide the identity operator $I_{4}$ into three parts $0.2I_{4},0.3I_{4},0.5I_{4}$ and construct a classical vector,
\begin{align}
\boldsymbol{y}=\frac{1}{\sqrt{\mathcal{N}_{y}}}[&\sqrt{0.2},\sqrt{0.3},\sqrt{0.5},\sqrt{11.814\mu},\sqrt{0.4366\mu},\nonumber\\
&\sqrt{12.25\mu},\sqrt{4.286\mu},\sqrt{4.286\mu}],
\end{align}
where the constant $\mathcal{N}_{y}=1+33.0726\mu$ \cite{Note2}.

\begin{figure}[ht]
\includegraphics[scale=0.55]{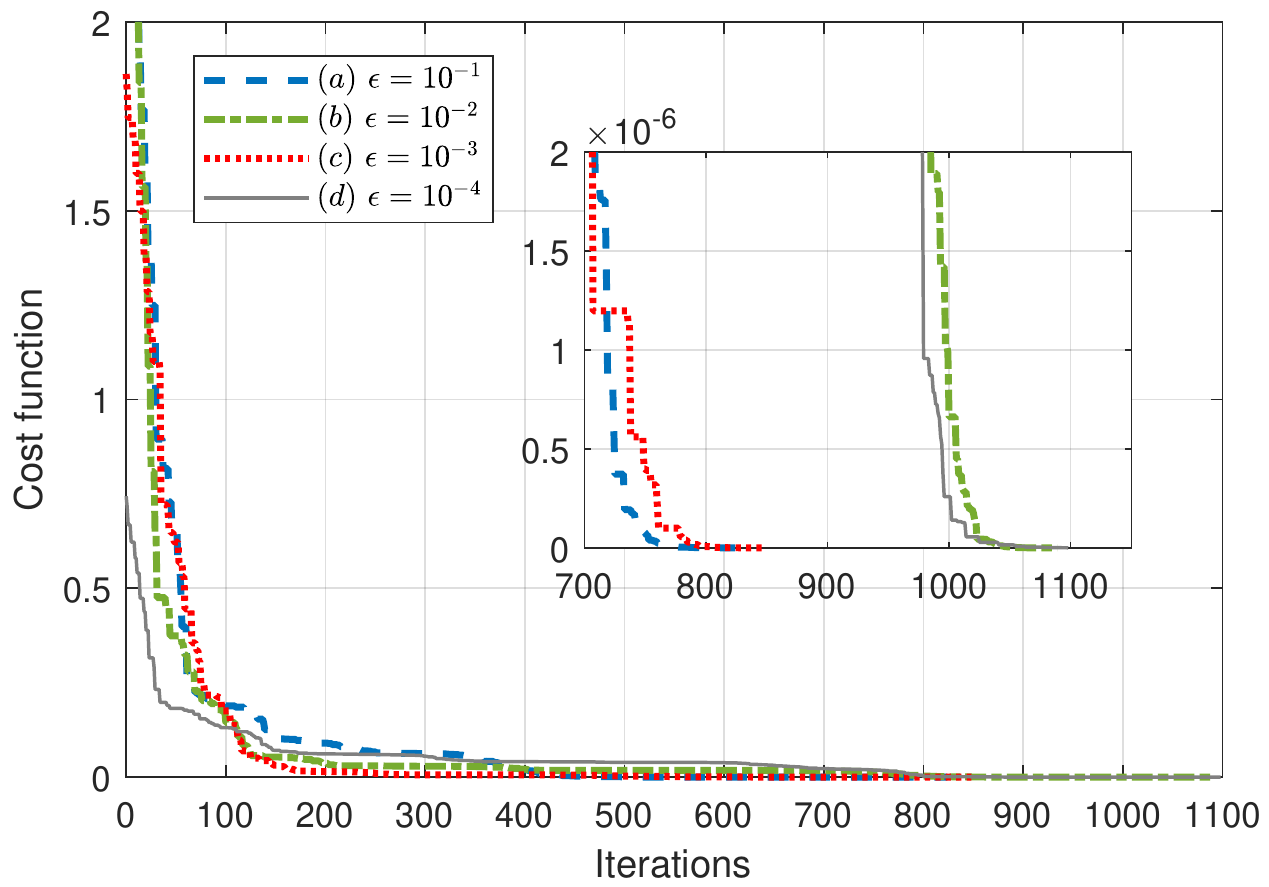}
\caption{The training processes of the cost function $F(\boldsymbol\theta)$ Eq. (\ref{CostFun}) under different learning rates. The final values the cost function are (a) $2.04389\times10^{-10}$ with $\epsilon=10^{-1}$, (b) $4.87097\times10^{-11}$ with $\epsilon=10^{-2}$, (c) $1.7075\times10^{-10}$  with $\epsilon=10^{-3}$ and (d) $3.02248\times10^{-10}$  with $\epsilon=10^{-4}$.}
\label{Fig2Training}
\end{figure}

The first step trains a PQC $U(\boldsymbol\theta)$ to approximately prepare a $3$-qubit quantum state $|\boldsymbol{y}\rangle\approx U(\boldsymbol\theta_{\textrm{opt}})|000\rangle$ which is an amplitude encoding state of the vector $\boldsymbol{y}$. In our numerical experiments, as shown in Fig. (\ref{Fig1schematic}.b), the circuit depth $L=3$ and $\boldsymbol\theta=(\theta_{1},\cdots,\theta_{9})$. Due to the relation between the learning rate $\mu$ and the precision $\epsilon$, $\mu=\sqrt{\epsilon}/2$, as discussed in Sec. \ref{Sec:IID:Equivalence}, the learning rate would increase when the precision increases. Fig. \ref{Fig2Training} illustrates the training processes of four cases with the precisions $\epsilon=10^{-1},10^{-2},10^{-3},10^{-4}$. The classical optimization procedure utilizes the Nelder-Mead algorithm \cite{Nelder1965A}. It is straightforward to check that the final value approaches the minimal value zero. Meanwhile, we verify that the fidelity between $|\boldsymbol{y}\rangle$ and $U(\boldsymbol\theta_{\textrm{opt}})|000\rangle$ approaches one in these four cases.

\begin{figure}[ht]
\includegraphics[scale=0.55]{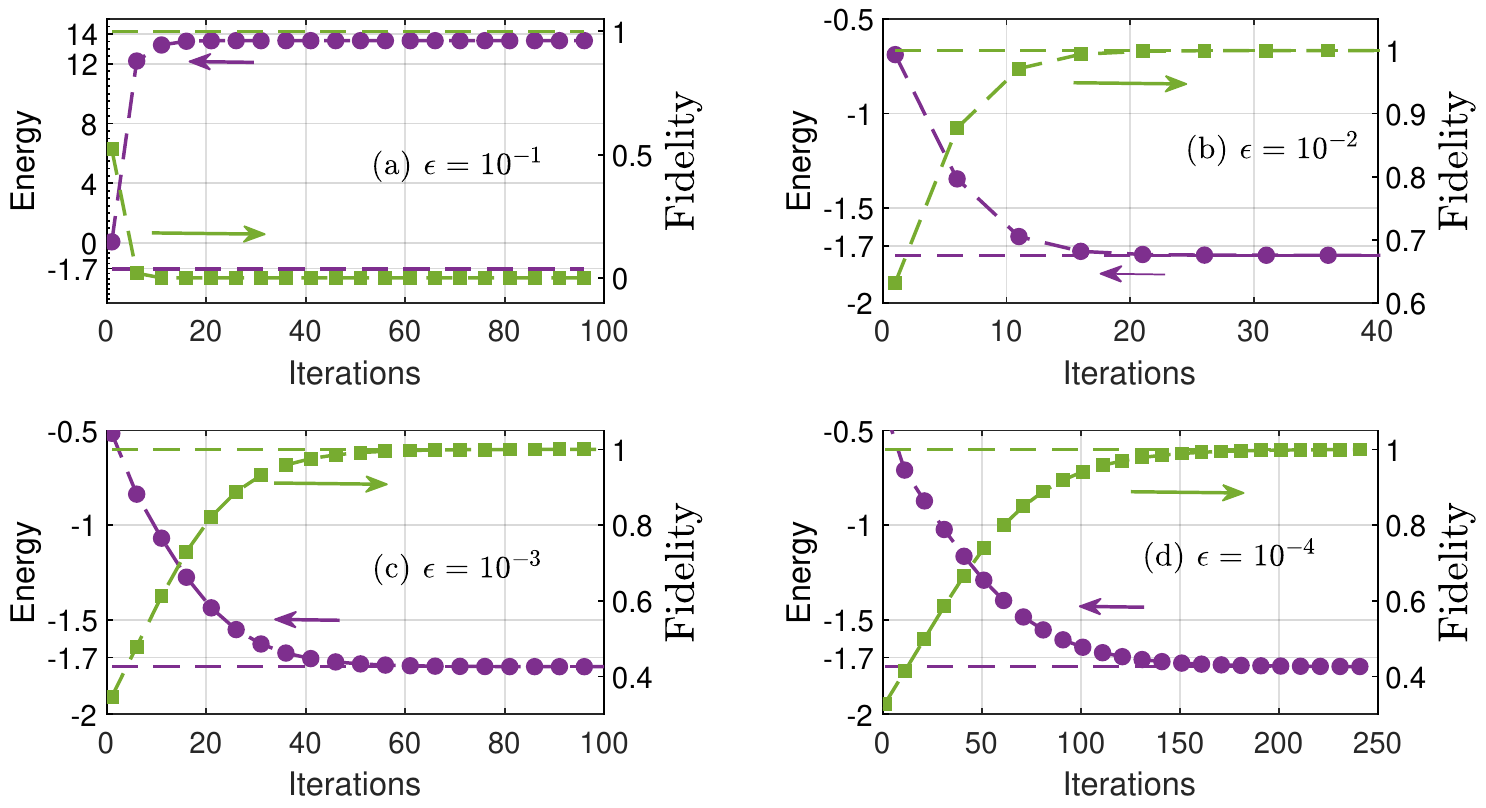}
\caption{The convergence curves of energy (left axis) and fidelity (right axis) for the deuteron molecule.}
\label{Fig3Result1}
\end{figure}

Next we iteratively apply the unitary $\Lambda_{\tilde{k}}G$ on an initial state $|\phi(0)\rangle=V(\boldsymbol\alpha)|00\rangle$, where
\begin{align}\label{InitialState}
V(\boldsymbol\alpha)=&[R_{y}(\alpha_1)R_{y}(\alpha_2)\otimes R_{y}(\alpha_3)R_{y}(\alpha_4)]\cdot\textrm{CNOT}
\end{align}
and the parameter $\boldsymbol\alpha=(0.3692,0.1112,0.7803,0.3897)$. The overlap $|c_{0}^{(0)}|^2=|\langle u_{0}|\phi(0)\rangle|^2=0.3186$. Fig. \ref{Fig3Result1} plots the numerical results under different precisions $\epsilon$. Notice that the energy does not converge to the ground state energy when $\epsilon=10^{-1}$ as shown in Fig. (\ref{Fig3Result1}.a). This situation means that the ITE is performed with a long time step $\Delta t=\sqrt{\epsilon}=\sqrt{10^{-1}}=0.3162$. Theoretically, in this case, the eigenvalues (in absolute value) of $G$ are
\begin{align}
&|\lambda_{0}|=|1-2\mu E_{0}|=1.5529,|\lambda_{1}|=|1-2\mu E_{1}|=0.9999,\nonumber\\
&|\lambda_{2}|=|1-2\mu E_{2}|=2.7358,|\lambda_{3}|=|1-2\mu E_{3}|=3.2889.\nonumber
\end{align}
Based on the power method \cite{Golub2013Matrix}, the algorithm converges to $E_{3}$ rather than the lowest eigenvalue $E_{0}$. When the precision $\epsilon\leq10^{-2}$, the energy and the fidelity can successfully converge to the exact value $-1.7485$ and $0.9999$ (see Fig. \ref{Fig3Result1}b,c,d). Note that, however, the smaller time step has the higher number of iteration step. In particular, about $200$ ($40$) steps are required to reach the ground state for the precision $\epsilon=10^{-4}$ ($10^{-2}$).

Until now we have focused on the performance of our algorithm under noiseless situation. As quantum devices are usually imperfect and have noise, we here consider the noise implementation by taking into account a global depolarizing noise $\mathcal{E}$ \cite{Schumacher1996Sending,Fontana2021Evaluating}. For any $n$-qubit state $\rho_{0}$, the global symmetric depolarizing noise channel is defined by
\begin{align}
\mathcal{E}(\rho_{0})=(1-\beta)\rho_{0}+\frac{\beta}{2^{n}}I_{2^{n}},
\end{align}
where the parameter $\beta$ denotes the strength of the noise. Table \ref{Comparison} shows the quantity $[E(s)-E_{0}]/|E_{0}|$ and the evolved fidelity $\textrm{Tr}(|u_{0}\rangle\langle u_0|\phi(s))$ between the exact ground state $|u_{0}\rangle$ and the iterated state $\phi(s)$ under different strength $\beta$. The fidelity is evaluated exactly for the state obtained in the noisy simulation. The corresponding calculation is just classical. Table \ref{Comparison} implies that the fidelity increases with the decreasing of $\beta$. It suggests that we still can obtain the ground state with amenable state fidelity even though the quantum device has a global noise.
\begin{table}
\caption{The numerical results of noisy simulation of our algorithm when the precision $\epsilon=10^{-2}$.}
\label{Comparison}
\begin{ruledtabular}
\begin{tabular}{c|cccccc}
 $\beta$ &0.00 & 0.02 & 0.04& 0.06& 0.08& 0.1 \\
 \hline
 $[E(s)-E_{0}]/|E_{0}|$ &0.0778 & 0.1272 & 0.1783& 0.2313& 0.2862 & 0.3431 \\
 \hline
 $\textrm{Tr}(|u_{0}\rangle\langle u_0|\phi(s))$ &0.9911 & 0.9805 & 0.9696& 0.9583& 0.9467& 0.9346  \\
\end{tabular}
\end{ruledtabular}
\end{table}

\begin{figure}[ht]
\includegraphics[scale=0.55]{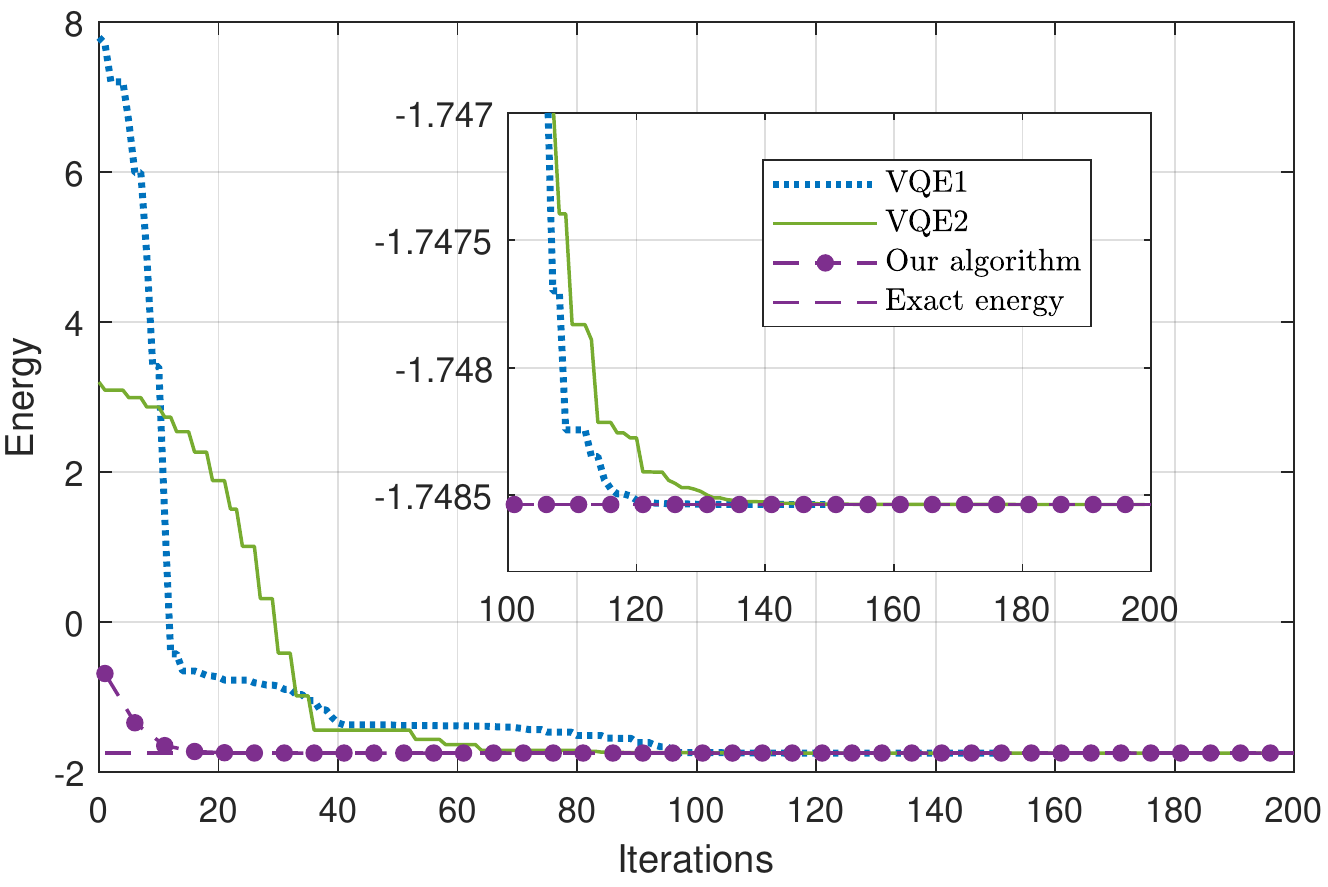}
\caption{The numerical simulation of VQE and our approach under conditions $\beta=0$ and $\epsilon=10^{-2}$. VQE1 and VQE2 denote the cases with the depth of the ansatz $1$ and $2$.}
\label{Fig4Result2}
\end{figure}

Furthermore, we utilize the VQE \cite{Peruzzo2014Variational} to obtain the ground state by minimizing the cost function $C(\boldsymbol\gamma)=\langle00|W^{\dag}(\boldsymbol\gamma)
\mathcal{H}W(\boldsymbol\gamma)|00\rangle$. The PQC is given by
\begin{align}
W(\boldsymbol\gamma)=&[R_{y}(\gamma_3)\otimes R_{y}(\gamma_{4})](|0\rangle\langle0|\otimes I_{2}+|1\rangle\langle1|\otimes\sigma_{z})\times\nonumber\\
&[R_{y}(\gamma_1)\otimes R_{y}(\gamma_{2})].\nonumber
\end{align}
We apply the Nelder-Mead algorithm to optimize the $4$ variational parameters and find the optimal parameter $\boldsymbol\gamma_{\textrm{opt}}=(0.2556,1.9895,-0.8996,-3.4769)$. We also consider a $2$ depth circuit with 8 parameters. As shown in Fig. \ref{Fig4Result2}, our result converges faster than the VQE whose successful implementation requires to choose suitable PQC, avoid the local minimal value and train the cost function, while our iteration quantum algorithm can always obtain the ground state without the so-called barren plateau.

\begin{figure}[ht]
\includegraphics[scale=0.5]{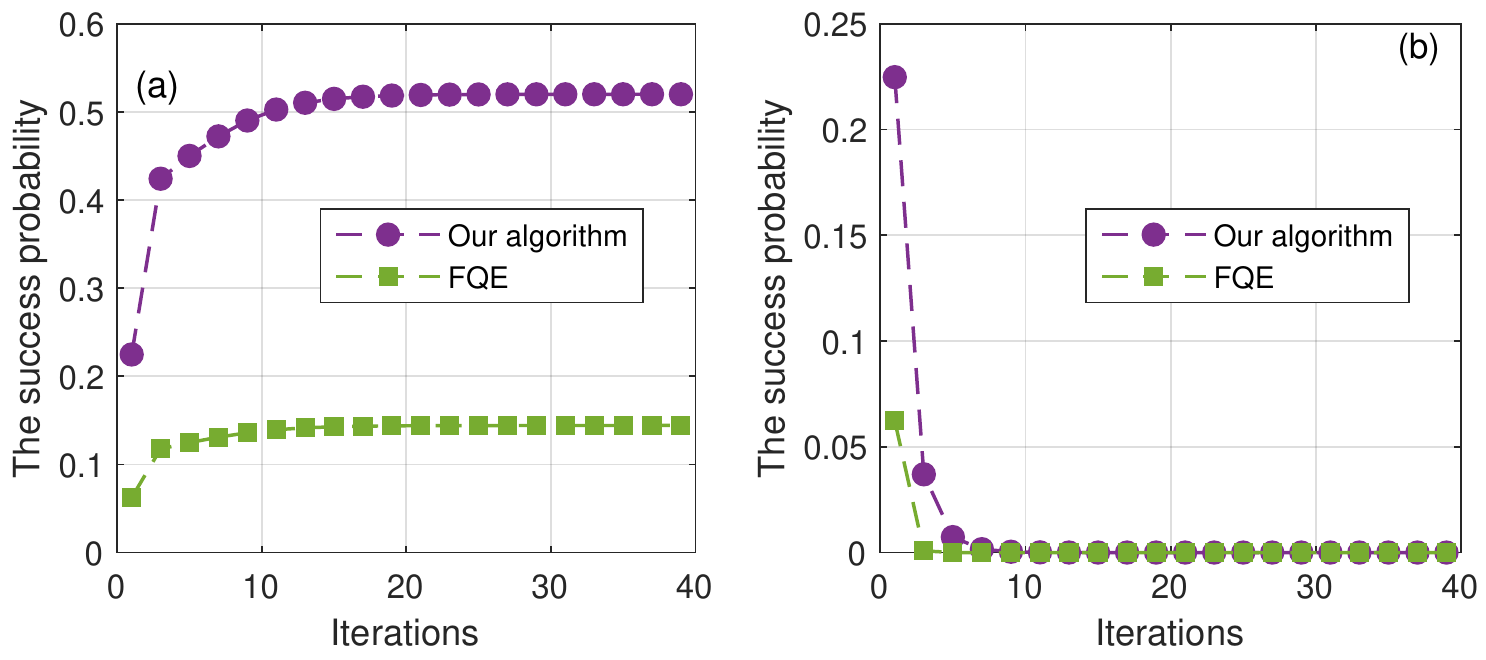}
\caption{(a) The \emph{local} success probability of each iteration. (b) The \emph{global} success probability of each iteration. In both cases, the learning rate $\mu=\sqrt{10^{-2}}/2$.}
\label{Fig5Result3}
\end{figure}

Finally, we investigate the success probability after each iteration. As shown in Fig. \ref{Fig5Result3} (a), our algorithm has higher success probability than FQE. This implies that our method requires much less repetition to prepare the iterated state and thus reduces the computation resources. Here, the success probability $P(s+1)$ corresponds to the measurement result when we have obtained the state $|\phi(s)\rangle$. In this case it can be seen as a \emph{local} success probability. If we consider the former $s$ steps as an overall procedure, the \emph{global} success probability is given by $P^{'}(s+1)=\prod_{j=1}^{s}P(j)$. Fig. \ref{Fig5Result3} (b) indicates that the \emph{global} success probability decays exponentially with the number of iterations.

\subsubsection{The Heisenberg model}

We now consider the following Heisenberg spin-1/2 chain Hamiltonian with open boundaries,
\begin{align}
\hat{\mathcal{H}}=&-J\sum_{j=1}^{n-1}(\sigma_{x}^{(j)}\sigma_{x}^{(j+1)}
+\sigma_{y}^{(j)}\sigma_{y}^{(j+1)}+\sigma_{z}^{(j)}\sigma_{z}^{(j+1)})\nonumber\\
&-h\sum_{j=1}^{n}\sigma_{z}^{(j)}.
\end{align}
for $n=2$ and $4$.

\begin{figure}[ht]
\includegraphics[scale=0.5]{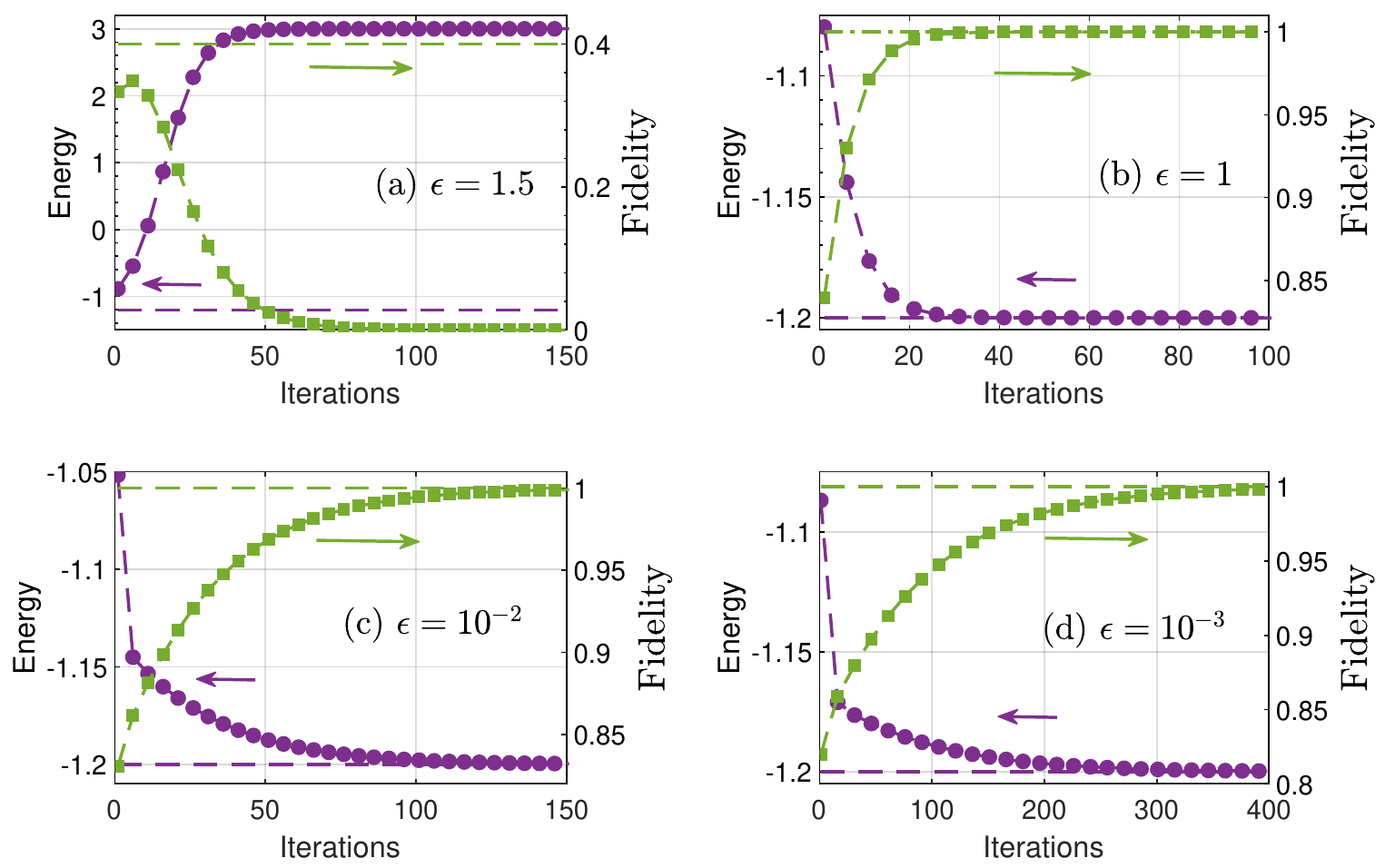}
\caption{The convergence curves of energy (left axis) and fidelity (right axis) for the Heisenberg model with $n=2$, $h=0.1$ and $J=1$. In VQSP, the trained values of the corresponding cost function $F(\boldsymbol{\theta}_{\textrm{opt}})$ are (a) $4.2\times10^{-13}$, (b) $4.132\times10^{-10}$, (c) $5.909\times10^{-10}$ and (d) $1.51\times10^{-12}$.}
\label{Fig6Result4}
\end{figure}

For $n=2$, the gradient operator is of the form,
\begin{align}
G&=I_{4}-2\eta\hat{\mathcal{H}}\nonumber\\
&=0.2I_{4}+0.4I_{4}+0.4I_{4}+2\mu J[\sigma_{x}^{(1)}\sigma_{x}^{(2)}+\sigma_{y}^{(1)}\sigma_{y}^{(2)}\nonumber\\
&+\sigma_{z}^{(1)}\sigma_{z}^{(2)}]+2\mu h(\sigma_{z}^{(1)}+\sigma_{z}^{(2)}).
\end{align}
Performing VQSP, we prepare the ancillary state $|\boldsymbol{y}\rangle$ associated with the vector
\begin{align}
\boldsymbol{y}=\frac{1}{\sqrt{\mathcal{N}_{y}}}[&\sqrt{0.2},\sqrt{0.4},\sqrt{0.4},\sqrt{2\mu J},\sqrt{2\mu J},\sqrt{2\mu J},\nonumber\\
&\sqrt{2\mu h},\sqrt{2\mu h}],
\end{align}
where the constant $\mathcal{N}_{y}=1+2\mu(3J+2h)$. The used PQC has $3$ layers and $9$ parameters. The structure is similar to Fig. \ref{Fig1schematic} (b). The initial state $|\phi(0)\rangle$ is produced by unitary (\ref{InitialState}) with same parameters. From Fig. \ref{Fig6Result4} (c) and (d), we see that the energy and fidelity convergence approaches the ground state and the exact value $1$ when the precision $\epsilon\leq10^{-2}$ $[\mu\in(0,\sqrt{0.01}/2=0.05)]$. For the Heisenberg model, the largest eigenvalue is $3$ and the smallest eigenvalue is $-1.2$. Based on Theorem 1, the theoretical interval of the learning rate is $\mu\in(0,\frac{1}{3-1.2}\approx0.5556)$ which contains the empirical interval $(0,0.05)$. There exists $\epsilon$ such that $\mu\in(0.05,0.5556)$ as shown in Fig. \ref{Fig6Result4} (b). When $\epsilon=1.5$, the learning rate $\mu=\sqrt{1.5}/2=0.6124\not\in(0,0.5556)$. Thus, in Fig. \ref{Fig6Result4} (a) the energy does not converge to the one of ground state.

\begin{figure}[ht]
\includegraphics[scale=0.5]{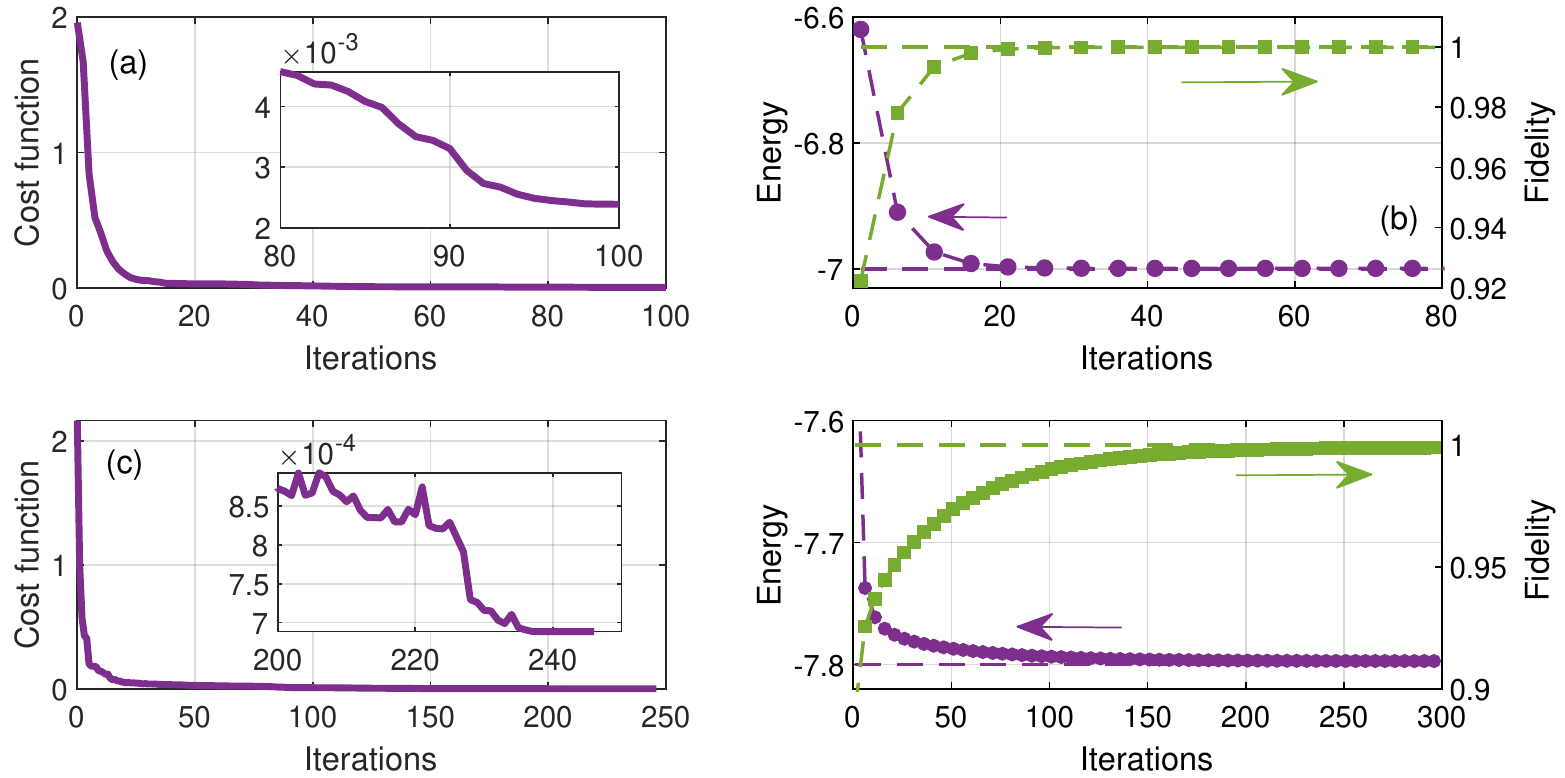}
\caption{(a) The convergence curves of energy (left axis) and fidelity (right axis) for the Heisenberg model with $n=8$ and $J=1$, $h=0.1$. The learning rate $\mu=\sqrt{10^{-2}}/2$. (b) The training process of the cost function $F(\boldsymbol{\theta})$. The final value is $2.38995\times10^{-3}$.}
\label{Fig7Result5}
\end{figure}

For the Heisenberg model with $n=4$ and $h=J=1$, the identity $I_{16}$ in the gradient operator is decomposed into three parts $0.2I_{16}+0.4I_{16}+0.4I_{16}$. For preparing the associated ancilla state $|\boldsymbol{y}\rangle$, the PQC has $2$ layers and each layer contains single rotations $R_y(\theta_i)$ and controlled rotations $R_y(\theta_j)$. Thus the number of parameters is $16$. The training of the cost function is shown in Fig. \ref{Fig7Result5} (a) and the final value is $2.3900\times10^{-3}$. The fidelity of preparing $|\boldsymbol{y}\rangle$ is $0.9992$. The initial state is $|\phi(0)\rangle=\bigotimes_{i=1}^{4}R_y(\alpha_i)|0^{\otimes4}\rangle$ with
\begin{align}
(\alpha_1,\alpha_2,\alpha_3,\alpha_4)=(0.5906,0.6604,0.0476,0.3488).
\end{align}
Fig. \ref{Fig7Result5} (b) shows that our algorithm can successful converge towards the ground energy.

For the Heisenberg model with $n=8$, $h=0.1$ and $J=1$, the identity $I_{256}$ in the gradient operator is decomposed into three parts $0.2I_{256}+0.4I_{256}+0.4I_{256}$. For preparing the associated ancilla state $|\boldsymbol{y}\rangle$, the PQC has $3$ layers and each layer contains single rotations $R_y(\theta_i)$ and controlled rotations $R_y(\theta_j)$. Thus the number of parameters is $30$. The training of the cost function is shown in Fig. \ref{Fig7Result5} (c) and the final value is $6.8840\times10^{-4}$. The fidelity of preparing $|\boldsymbol{y}\rangle$ is $0.9998$. The initial state is $|\phi(0)\rangle=\bigotimes_{i=1}^{8}R_y(\alpha_i)|0^{\otimes8}\rangle$ with
\begin{align}
(\alpha_1,\cdots,\alpha_8)=(&0.1079,0.1822,0.0991,0.4898,0.1932,\nonumber\\&
0.8959,0.0991,0.0442).
\end{align}
Fig. \ref{Fig7Result5} (d) shows that our algorithm can successful converge towards the ground energy.

\subsection{Performance and scaling of different algorithms}
In this subsection, we analyse and compare the computational complexity of our approach with other ground state estimation methods. The computational complexity contains the number of qubits (qubit complexity), the number of quantum gates (gate complexity), and the number of measurements (sampling complexity).

First of all, we evaluate the computational complexity of our iterative algorithm. The number of qubits used to store the state $|\boldsymbol{y}\rangle|\phi(s)\rangle$ in two registers is $n+\tilde{k}$. Our algorithm includes the VQSP part and the block unitary evolution part. The gate complexity of VQSP scales as $\mathcal{O}[\textrm{poly}(\tilde{k})]$ with $\tilde{k}=\log_{2}(K+1)$, where the constant $K$ scales polynomial on the number of qubits as $\mathcal{O}[\textrm{poly}(n)]$. Thus the overhead of VQSP is $\mathcal{O}[\textrm{poly}(\log n)]$. As analysed in \ref{Sec:IIB:Unitary}, the cost of performing unitary part scales as $\mathcal{T}_{\textrm{total}}=\mathcal{O}[l(K+1)\widetilde{k}^{2}]=\mathcal{O}[l\textrm{poly}(n)(\log n)^{2}]\approx\mathcal{O}[l\textrm{poly}(n)]$. Here we neglect the item $(\log n)^{2}$. As a result, the gate complexity of our algorithm is
\begin{align}\label{ComplexityA}
\mathcal{O}[lN_{t}\textrm{poly}(n)+2N_{t}\textrm{poly}(\log n)],
\end{align}
which depends polynomially on the number of Trotter steps $N_{t}$, qubit number $n$ and locality $l$. For an $n$-qubit and $l$-local Hamiltonian $\hat{\mathcal{H}}$, a classical computer to perform imaginary time evolution in general requires time and space that scale at least as $\textrm{exp}[\mathcal{O}(2^{n})]$. Our algorithm overcomes these bottlenecks and reduces the runtime to $\mathcal{O}[lN_{t}\textrm{poly}(n)]$ which exhibits an exponential speedup in the size of the system compared with classical methods. At each iteration, the sample complexity determined by the success probability scales as $\mathcal{O}[P(s+1)^{-1}]$. According to the property of the success probability, the number of measurements is
\begin{align}
\frac{1}{P(1)}=\frac{\mathcal{N}_{y}^{2}}{\|G|\phi(0)\rangle\|_{2}^{2}}
=\frac{\sum_{i=0}^{N-1}y_k}{\sum_{i=0}^{N-1}|\lambda_i|^{2}|c_i^{(0)}|^2}
<\frac{\sum_{i=0}^{N-1}y_k}{|\lambda_0|^{2}|c_0^{(0)}|^2}.\nonumber
\end{align}
Notice that the sampling complexity is related to three quantities including the summation of coefficients $\sum_{i=0}^{K}y_{k}$, the overlap $c_0^{(0)}$ between the initial state $|\phi(0)\rangle$ and the ground state $|u_0\rangle$, and the largest (in absolute value) finite eigenvalue $|\lambda_0|$ of the gradient operator $G$. Let the summation of coefficients $\sum_{i=0}^{K}y_{k}\sim\mathcal{O}[\textrm{poly}(n)]$ scales polynomially in the number of qubits and the overlap $|c_0^{(0)}|^2\sim\mathcal{O}[1/\textrm{poly}(n)]$. In this case, the sampling complexity scales $\mathcal{O}[\textrm{poly}(n)]$ which increases polynomially with the number of qubits, implying that our approach is practical for estimating the ground states. To decrease the sample complexity of an approach is clearly to increase $|c_{0}^{(0)}|^2$ by using a variational method \cite{Note1}.

A promising approach for preparing a ground state of a Hamiltonian on near-term devices is the variational quantum eigensolver (VQE) with hybrid quantum-classical loops \cite{Peruzzo2014Variational}. The classical computer trains a PQC with shallow depth and learns an optimal parameter which is fed into an NISQ device to produce the ground state \cite{McClean2016The}. However, the efficiency of VQE highly depends on the ansatz choice \cite{Patti2021Entanglement} and the optimization of the non-convex cost function \cite{Wang2021Noise}. The classical optimization problems are generally NP-hard and the gradient descent algorithms may not converge to the optimal solution \cite{Bittel2021Training}. Due to the fact that our iterative algorithm is based on the ITE, it circumvents the aforementioned potential problems if the subroutine VQSP is successful. It is worth to remark that our algorithm prepares the ground state on the original state space rather than parameter space, namely, it can always reach the global minimal. Moreover, the VQE and our algorithm consume different quantum resources. For a single step, the VQE only involves a trial state preparation which has an overhead $\mathcal{O}(\textrm{poly}(n))$ \cite{Peruzzo2014Variational}. Based on Eq. (\ref{ComplexityA}), the overhead of our algorithm is $\mathcal{O}[l\textrm{poly}(n)+2\textrm{poly}(\log n)]$. Thus, the VQE has small gate cost compared with our algorithm in this case.

Another iterative quantum algorithm based on LCU is the FQE \cite{Wei2020A}. FQE starts with a different ancillary state,
\begin{align}
|\tilde{\boldsymbol{y}}\rangle=\frac{1}{\sqrt{\widetilde{\mathcal{N}}_{y}}}\sum_{k=0}^{K}y_{k}|k\rangle,
~~\widetilde{\mathcal{N}}_{y}=\sum_{k=0}^{K}y_{k}^2.
\end{align}
After performing the controlled unitary $\Lambda_{\tilde{k}}G$ on the composed state $|\tilde{\boldsymbol{y}}\rangle\otimes|\phi(s)\rangle$, one obtains the state
\begin{align}
\frac{1}{\sqrt{\widetilde{\mathcal{N}}_{y}}}\sum_{k=0}^{K}y_{k}|k\rangle\otimes G_{k}|\phi(s)\rangle.
\end{align}
Next one applies $\tilde{k}$ Hadamard gate $H$ on the ancillary system and gets
\begin{align}
\frac{1}{\sqrt{(K+1)\widetilde{\mathcal{N}}_{y}}}
\sum_{k,j=0}^{K}y_{k}(-1)^{k\cdot j}|j\rangle\otimes G_{k}|\phi(s)\rangle,
\end{align}
where $k\cdot j$ denotes the bitwise inner product of $k$ and $j$ modulo 2. Notice that this step is different from our approach. In our algorithm, we apply the unitary $U^{\dag}(\boldsymbol\theta_{\textrm{opt}})$ rather than Hadamard gates. Using the projector operator $M$ to measure the above state, one gets the collapsed state,
\begin{align}
\frac{1}{\sqrt{\widetilde{P}(s)(K+1)
\widetilde{\mathcal{N}}_{y}}}|0^{\otimes\tilde{k}}\rangle
\sum_{k=0}^{K}y_{k}G_{k}|\phi(s)\rangle,
\end{align}
which is proportional to state $|\phi(s+1)\rangle=\frac{G|\phi(s)\rangle}{\|G|\phi(s)\rangle\|_{2}}$. The success probability of obtaining the result $|0^{\otimes\tilde{k}}\rangle$ is
\begin{align}
\widetilde{P}(s)=\frac{\|G|\phi(s)\rangle\|_{2}^{2}}{(K+1)\widetilde{\mathcal{N}}_{y}}
=\frac{\|G|\phi(s)\rangle\|_{2}^{2}}{(K+1)\sum_{k=0}^{K}y_{k}^2}.
\end{align}

\begin{lemma}\label{Lemma2}
The success probability $\widetilde{P}(s)$ is an increasing sequence such that
\begin{align}
\widetilde{P}(1)\leq\widetilde{P}(2)\leq\cdots\leq\widetilde{P}(S).
\end{align}
\end{lemma}
The proof of Lemma 2 is similar the proof of Lemma 1. Due to the fact that
\begin{align}
\Bigg[\sum_{k=0}^{K}y_{k}\Bigg]^2-(K+1)\sum_{k=0}^{K}y_{k}^2&=-\sum_{k=0}^{K}\sum_{j=k+1}^{K}(y_k-y_j)^2\leq0,\nonumber
\end{align}
we obtain $\mathcal{N}_{y}^{2}\leq(K+1)\widetilde{\mathcal{N}}_{y}$. Thus, the success probability of our approach is higher than the counterpart of the FQE, $P(s)\geq\widetilde{P}(s)$. This property indicates that the sampling complexity of our algorithm is smaller than that of FQE.

As the gate complexity of preparing ancillary state is $\mathcal{O}(2^{\tilde{k}})=\mathcal{O}[\textrm{poly}(n)]$, the total gate complexity of FQE is
\begin{align}\label{ComplexityB}
\mathcal{O}[lN_{t}\textrm{poly}(n)+N_{t}\textrm{poly}(n)].
\end{align}
Compared with Eq. (\ref{ComplexityA}), it is clear that our algorithm reduces the gate complexity. However, the circuit depth of FQE and our algorithm are $D_{1}+D_{2}$ and $2D_{1}+D_{2}$, respectively, where $D_1$ and $D_2$ denote the depth of VQSP and the block diagonal unitary. Thus FQE has more shallower circuit depth than our algorithm in this case.

Concerning the training of the cost function Eq. (\ref{CostFun}), there could be multiple local minima due to the selected ansatz. It is a key problem to find the global minimal value. Similar to general VQAs, our algorithm also encounters the so-called barren plateau (BP) phenomenon \cite{Cerezo2021Variational}. In our numerical simulations, we have employed different initial points until the global minimal is reached. However, several other approaches may also be used in our algorithm such as designing local cost function \cite{Cerezo2021Cost} and constructing adaptive circuit structure \cite{Pesah2021Absence}.

\section{Comparison with quantum imaginary time evolution}
We remark that our algorithm framework can be treated as a quantum version of ITE. In this section we give a detailed comparison with the quantum imaginary time evolution (QITE) \cite{Motta2020Determining}.

Given a Hamiltonian $\hat{\mathcal{H}}=\sum_{k=0}^{K-1}h_k\hat{\mathcal{H}}_k$ and an initial state $|\phi\rangle$. For a small step $\Delta t$, the Trotter-Suzuki decomposition can simulate the evolution,
\begin{align}
e^{-\Delta t\hat{\mathcal{H}}}\approx e^{-\Delta t\hat{\mathcal{H}}_{1}}e^{-\Delta t\hat{\mathcal{H}}_{2}}\cdots e^{-\Delta t\hat{\mathcal{H}}_{K}},
\end{align}
where the Trotter error subsumes terms of order $\Delta t^2$ and higher. The QITE replaces each non-unitary step $e^{-\Delta t\hat{\mathcal{H}}_{k}}$ with a unitary evolution $e^{-\iota\Delta t\hat{A}_k}$ \cite{Motta2020Determining}. More specially, the goal of QITE is to minimize the difference between states
\begin{align}
|\phi^{'}\rangle&=\frac{e^{-\Delta t\hat{\mathcal{H}}_k}|\phi\rangle}{\sqrt{\langle\phi|e^{-2\Delta t\hat{\mathcal{H}}_k}|\phi\rangle}}
\approx\frac{(I_N-\Delta t\hat{\mathcal{H}}_k)|\phi\rangle}{\sqrt{\langle\phi|(I_N-\Delta t\hat{\mathcal{H}}_k)|\phi\rangle}},
\end{align}
and
\begin{align}
e^{-\iota\Delta t\hat{A}_k}|\phi\rangle\approx(I_N-\iota\Delta t\hat{A}_k)|\phi\rangle.
\end{align}
Decomposing the $D$-qubit operator $\hat{A}_k$ as a sum of Pauli strings, we have
\begin{align}
\hat{A}_k=\sum_{i_1i_2\cdots i_D}a[k]_{i_1i_2\cdots i_D}\sigma_{i_1}\sigma_{i_2}\cdots\sigma_{i_D}
=\sum_{\mathcal{I}}a[k]_{\mathcal{I}}\sigma_{\mathcal{I}},\nonumber
\end{align}
where $\mathcal{I}$ denotes the index $i_1i_2\cdots i_D$. The minimum value of the function
\begin{align}
&\left\|\frac{|\phi^{'}\rangle-|\phi\rangle}{\Delta t}+\iota\hat{A}_k|\phi\rangle\right\|_2
\end{align}
can be approximately obtained by solving the linear equation $(\boldsymbol{S}+\boldsymbol{S}^{T})\boldsymbol{a}=-\boldsymbol{b}$, where $\boldsymbol{S}$ and $\boldsymbol{b}$ are the expectation values obtained by measuring $|\phi\rangle$.

Different from QITE, our algorithm achieves the non-unitary $I_N-\Delta t\hat{\mathcal{H}}_k$ by adding a single ancilla qubit. By applying a unitary
\begin{align}
U=\begin{pmatrix}
\frac{1}{\sqrt{1+\Delta t}} & \frac{\sqrt{\Delta t}}{\sqrt{1+\Delta t}}\\
\frac{-\sqrt{\Delta t}}{\sqrt{1+\Delta t}} & \frac{1}{\sqrt{1+\Delta t}}
\end{pmatrix}=R_y(\boldsymbol{\theta}_{\textrm{opt}})
\end{align}
on state $|0\rangle$, we prepare a superposition state $|\boldsymbol{y}\rangle=\frac{1}{\sqrt{1+\Delta t}}|0\rangle-\frac{\sqrt{\Delta t}}{\sqrt{1+\Delta t}}|1\rangle$, where the parameter $\boldsymbol{\theta}_{\textrm{opt}}=2\arccos\frac{1}{\sqrt{1+\Delta t}}$. Next, we apply the unitary $U^{\dag}(|0\rangle\langle0|\otimes I_N+|1\rangle\langle1|\otimes\hat{\mathcal{H}}_k)$ to the state $|\boldsymbol{y}\rangle\otimes|\phi\rangle$, which yields the following state,
\begin{align}
&|0\rangle\left(\frac{1}{1+\Delta t}|\phi\rangle-\frac{\Delta t}{1+\Delta t}\hat{\mathcal{H}}_k|\phi\rangle\right)-\nonumber\\
&|1\rangle\left(\frac{\sqrt{\Delta t}}{1+\Delta t}|\phi\rangle-\frac{\sqrt{\Delta t}}{1+\Delta t}\hat{\mathcal{H}}_k|\phi\rangle\right).
\end{align}
Finally, we measure the ancilla qubit to get the resulting state $|0\rangle$. The success probability is
\begin{align}
P&=\frac{1}{\left(1+\Delta t\right)^2}
\left\||\phi\rangle-\Delta t\hat{\mathcal{H}}_k|\phi\rangle\right\|_2^2,
\end{align}
and the collapsed state is
\begin{align}
|\phi^{'}\rangle=\frac{|\phi\rangle-\Delta t\hat{\mathcal{H}}_k|\phi\rangle}{\left(1+\Delta t\right)\sqrt{P}}
=\frac{(I_N-\Delta t\hat{\mathcal{H}}_k)|\phi\rangle}{\left\|(I_N-\Delta t\hat{\mathcal{H}}_k)|\phi\rangle\right\|_2}.
\end{align}

In summary, our algorithm requires one single ancilla qubit to achieve the imaginary time evolution while the QITE needs no any other ancilla qubit. Moreover, our approach is a totally quantum procedure. It does not use any classical part compared with QITE. Based on the locality of the Hamiltonian, the controlled unitary can be decomposed into $l$ single controlled Pauli operators. More importantly, all the required quantum gates are two qubit gates. Therefore, our method is suitable for NISQ devices.

\section{Conclusion and discussion}
The selection of learning rate and the preparation of ancillary state are two crucial problems for an iterative quantum algorithm to estimate the ground state of quantum systems. We have established an equivalence between imaginary time evolution (ITE) and our improved iterative quantum algorithm. This equivalence can be thought of to be a benchmark to determine the learning rate in terms of the time step of the ITE. Furthermore, we have utilized a variational quantum algorithm to produce an amplitude encoding quantum state whose elements are associated with the coefficients of the gradient operator. This preparation strategy requires polynomial resources in contrast with other preparation approaches, for instance, the Grover-oracle-based method \cite{Soklakov2006Efficient}, the decomposition of universal quantum gate \cite{Plesch2011Quantum} and ancillary-assisted methods \cite{Zhang2021Low,Zhang2022Quantum}. This crucial subroutine enables our algorithm to have lower gate complexity. Notice that Nakaji \emph{et.al.} proposed an efficient method for approximate amplitude encoding by using a shallow PQC \cite{Nakaji2022Approximate}. Their method is more general and tackles the real-valued vector rather than positive vector. However, the defined cost function is different. The numerical results demonstrate that our proposal successfully finds the ground state of the deuteron molecule with high state fidelity. Our approach not only estimates the ground state energy but also prepares the ground state.

Our approach has general applicability to problems in many-body physics,
since it is potential to implement a non-unitary operator on an arbitrary state.
For example, given the Lindblad operator of open quantum systems,
the entire non-unitary evolution can be simulated by combining the vectorizing
density matrix with our approach. The iteration quantum framework can naturally
be used to prepare non-equilibrium steady states \cite{Liang2022Quantum},
excited states \cite{Wen2021A} and to compute the time domain Green's
function \cite{Keen2021Quantum}.

\bigskip

{\sf Acknowledgements:} This work is supported by the National Natural Science Foundation of China (NSFC) under Grants 12075159 and 12171044; Beijing Natural Science Foundation (Grant No. Z190005); Academy for Multidisciplinary Studies, Capital Normal University; the Academician Innovation Platform of Hainan Province; Shenzhen Institute for Quantum Science and Engineering, Southern University of Science and Technology (No. SIQSE202001); Shandong Provincial Natural Science Foundation for Quantum Science No.ZR2020LLZ003, ZR2021LLZ002, and the Fundamental Research Funds for the Central Universities No.22CX03005A.

{\sf Conflict of Interest:} The authors declare no conflict of interest.

{\sf Data Availability Statement:} The data that support the findings of this study are available from the corresponding author upon reasonable request

\begin{appendix}
\section{Proof of the Lemma 1}\label{AppendixA}
As shown in the main text, the success probability of each iteration is
\begin{align}\label{SuccPro1}
P(s+1)=\frac{\|G|\phi(s)\rangle\|_2^2}{\mathcal{N}_{y}^2}.
\end{align}
After making a measurement, we get the state $|\phi(s+1)\rangle$ with probability $P(s+1)$. Expanding $|\phi(s)\rangle$ in the eigenbasis $\{|u_i\rangle\}_{i=0}^{N-1}$, we obtain
\begin{align}\label{DecomA}
|\phi(s)\rangle=\sum_{i=0}^{N-1}c_{i}^{(s)}|u_i\rangle,~~\sum_{i=0}^{N-1}|c_{i}^{(s)}|^2=1.
\end{align}
Plugging Eq. (\ref{DecomA}) in Eq. (\ref{SuccPro1}), we have
\begin{align}\label{SuccPro2}
P(s+1)=\frac{\sum_{i=0}^{N-1}|\lambda_i|^2|c_{i}^{(s)}|^2}{\mathcal{N}_{y}^2}.
\end{align}
By the same trick, the success probability of obtaining state $|\phi(s+2)\rangle$ is given by
\begin{align}\label{SuccPro3}
P(s+2)=\frac{\sum_{i=0}^{N-1}|\lambda_i|^2|c_{i}^{(s+1)}|^2}{\mathcal{N}_{y}^2}.
\end{align}
Here, the state $|\phi(s+1)\rangle$ is given by
\begin{align}
|\phi(s+1)\rangle&=\frac{G|\phi(s)\rangle}{\|G|\phi(s)\rangle\|_2}
=\frac{\sum_{i=0}^{N-1}\lambda_ic_{i}^{(s)}|u_i\rangle}
{\sqrt{\sum_{i=0}^{N-1}|\lambda_i|^2|c_{i}^{(s)}|^2}}\nonumber\\
&=\sum_{i=0}^{N-1}c_{i}^{(s+1)}|u_i\rangle,
\end{align}
where the coefficients
\begin{align}
c_{i}^{(s+1)}=\frac{\lambda_ic_{i}^{(s)}}{\sqrt{\sum_{i=0}^{N-1}|\lambda_i|^2|c_{i}^{(s)}|^2}}.
\end{align}

We now show $P(s+2)\geq P(s+1)$ by showing $P(s+2)/P(s+1)\geq1$. It is clear to see that
\begin{align}
\frac{P(s+2)}{P(s+1)}&=\frac{\sum_{i=0}^{N-1}|\lambda_i|^2|c_{i}^{(s+1)}|^2}
{\sum_{i=0}^{N-1}|\lambda_i|^2|c_{i}^{(s)}|^2}
=\frac{\sum_{i=0}^{N-1}|\lambda_i|^4|c_{i}^{(s)}|^2}
{\Big[\sum_{i=0}^{N-1}|\lambda_i|^2|c_{i}^{(s)}|^2\Big]^2}.\nonumber
\end{align}
By using the inequality $|\lambda_i|\geq|\lambda_j|$ for $i\leq j$ and the condition $\sum_{i=0}^{N-1}|c_i^{(s)}|^2=1$, we have
\begin{align}
&\sum_{i=0}^{N-1}|\lambda_i|^4|c_{i}^{(s)}|^2
-\Bigg[\sum_{i=0}^{N-1}|\lambda_i|^2|c_{i}^{(s)}|^2\Bigg]^2\nonumber\\
&=\sum_{i,j=0}^{N-1}|\lambda_i|^4|c_{i}^{(s)}|^2|c_{j}^{(s)}|^2
-\sum_{i,j=0}^{N-1}|\lambda_i|^2|c_{i}^{(s)}|^2|\lambda_j|^2|c_{j}^{(s)}|^2\nonumber\\
&=\sum_{i,j=0}^{N-1}|\lambda_i|^2(|\lambda_i|^2-|\lambda_j|^2)|c_{i}^{(s)}|^2|c_{j}^{(s)}|^2\nonumber\\
&=\sum_{i=0}^{N-1}\sum_{j=i+1}^{N-1}\left(|\lambda_i^2|-|\lambda_j^2|\right)^2|c_{i}^{(s)}|^2|c_{j}^{(s)}|^2\nonumber\\
&\geq0.
\end{align}
Therefore, we obtain $P(s+2)/P(s+1)\geq1$.

\section{Proof of Theorem 1}\label{AppendixB}
We here prove the Theorem 1. Consider the eigenvalue decomposition of the non-unitary operator $G=\sum_{i=0}^{N-1}\lambda_{i}|u_i\rangle\langle u_i|$, $\lambda_{i}=1-2E_i\mu$. We define functions $W_{i}(\mu)=|\lambda_{i}|=|1-2E_i\mu|$ with variable $\mu$ for $i=0,\cdots,N-1$. Based on the power method \cite{Golub2013Matrix}, the evolved state $|\phi(t)\rangle$ converges to the ground state of $\hat{\mathcal{H}}$ if
\begin{align}\label{ConditionA}
|W_{0}(\mu)|>|W_{1}(\mu)|\geq\cdots\geq|W_{N-1}(\mu)|.
\end{align}
Let us discuss the upper bound of the learning rate $\mu$ according to following three cases.

Case 1: $0\leq E_0<E_1\leq\cdots\leq E_{N-1}$.

Case 2: $E_0<E_1\leq\cdots\leq E_{N-1}\leq0$.

Case 3: $E_0<E_1\leq \cdots\leq E_{i-1}\leq0\leq E_{i+1}\leq\cdots\leq E_{N-1}$.

In Case 1, we can verify that $W_{i}(\mu)\in[0,1]$ in the interval $\mu\in[0,\frac{1}{E_i}]$ and is a symmetric function about the axis $\mu=\frac{1}{2E_i}$ for all $i$. For $\mu\in(\frac{1}{2E_0},+\infty)$, the following equality always hold,
\begin{align}
W_{N-1}(\mu)\geq\cdots\geq W_{1}(\mu)>W_{0}(\mu),
\end{align}
which does not satisfy the condition (\ref{ConditionA}). In order to ensure the condition (\ref{ConditionA}) for $\mu\in(0,\frac{1}{2E_0})$, we set $W_i(\mu)=W_0(\mu)$ and obtain $\mu=\frac{1}{E_i+E_0}$. Accounting to the inequality $\frac{1}{E_{N-1}+E_0}\leq\cdots\leq\frac{1}{E_1+E_0}$, one sees that the algorithm converges to the ground state if $\mu\in(0,\frac{1}{E_{N-1}+E_0})$. In Case 2, we have an increasing sequence $|E_0|>|E_1|\geq\cdots\geq|E_{N-1}|\geq0$. Hence, the condition (\ref{ConditionA}) is always true for $\mu\in(0,+\infty)$. In Case 3, we have an inequality $|E_0|>|E_1|\geq\cdots\geq|E_{i-1}|\geq0$ and further
\begin{align}
1+2|E_0|\mu&>1+2|E_1|\mu\geq\cdots\geq1+2|E_{i-1}|\mu\geq1\nonumber
\end{align}
for $\mu>0$. In order to maintain the sequence ($\ref{ConditionA}$), we let $W_{N-1}(\mu)<1+2|E_0|\mu$ and find $2(E_{N-1}+E_0)\mu<1$. Notice that if $0\leq E_{N-1}\leq|E_0|$ $(E_0+E_{N-1}\leq0)$, the inequality (\ref{ConditionA}) always holds for $\mu>0$. For $E_{N-1}>|E_0|$ $(E_0+E_{N-1}>0)$, we obtain an upper bound of $\mu$, $\mu\in(0,\frac{1}{E_{N-1}+E_0})$. From the above analysis, we complete the proof.

\end{appendix}

\end{document}